\documentclass[prd,aps,onecolumn,nofootinbib]{revtex4}

\usepackage{enumitem}
\usepackage{graphicx}
\usepackage{subfigure}
\usepackage{amssymb}
\usepackage{amsmath}
\usepackage{bbm}
\usepackage{color}

\newcommand{\ignore}[1]{}
\newcommand{\be}{\begin{equation}} \newcommand{\ee}{\end{equation}}
\newcommand{\bea}{\begin{eqnarray}} \newcommand{\eea}{\end{eqnarray}}

\begin{document}

\title{Alignments in quasar polarizations: \\ pseudoscalar-photon mixing in the presence of correlated magnetic fields}
\author{Nishant Agarwal$^1$}
\author{Archana Kamal$^2$}
\author{Pankaj Jain$^3$}

\affiliation{$^1$Department of Astronomy, Cornell University, Ithaca, New York - 14853, USA \\
$^2$Department of Physics, Yale University, New Haven, Connecticut - 06520, USA \\
$^3$Department of Physics, Indian Institute of Technology, Kanpur - 208016, India}

\begin{abstract}
We investigate the effects of pseudoscalar-photon mixing on electromagnetic radiation in the presence of correlated extragalactic magnetic fields. We model the Universe as a collection of magnetic domains and study the propagation of radiation through them. This leads to correlations between Stokes parameters over large scales and consistently explains the observed large-scale alignment of quasar polarizations at different redshifts within the framework of the big bang model.
\end{abstract}

\maketitle


\section{Introduction}
\label{one}

Light pseudoscalar particles, such as the axion, arise naturally as pseudo-Goldstone bosons in spontaneously broken global symmetries \cite{Peccei:1977hh,Peccei:1977ur,Weinberg:1977ma,Wilczek:1977pj,McKay:1977gd,McKay:1978wn,Kim:1979if,Dine:1981rt,Kim:1986ax}. An axion has an effective coupling to two photons, therefore in an external magnetic field it can oscillate into a photon and vice versa \cite{Clarke:1982,Sikivie:1983ip,Sikivie:1985yu,Sikivie:1988mz,Maiani:1986md,Raffelt:1987im,Carlson:1994yqa,Bradley:2003kg,Das:2004qka,Das:2004ee,Ganguly:2005se,Ganguly:2008kh}. This mixing between axions and photons can lead to observable changes in the intensity and polarization of photons, having interesting astrophysical and cosmological implications \cite{Harari:1992ea,Berezhiani,Mohanty:1993nh,Das:2000ph,Kar:2000ct,Kar:2001eb,Csaki:2001jk,Csaki:2001yk,Grossman:2002by,Jain:2002vx,Song:2005af,Mirizzi:2005ng,Raffelt:2006cw,Gnedin:2006fq,Mirizzi:2007hr,Finelli:2008jv}. Since the mixing depends on the frequency of radiation, it also affects the electromagnetic spectrum \cite{Ostman:2004eh,Lai:2006af,Hooper:2007bq,Hochmuth:2007hk,Chelouche:2008ta}. Various experimental searches of pseudoscalar particles have led to significant limits on their masses and on the coupling parameters of the models
\cite{Dicus:1978fp,Vysotsskii:1978,Dearborn:1985gp,Raffelt:1987yu,Raffelt:1987yt,Turner:1987by,Mohanty:1993nh,Janka:1995ir,Keil:1996ju,Brockway:1996yr,Grifols:1996id,Raffelt:1999tx,Rosenberg:2000wb,Zioutas:2004hi,Yao:2006px,Raffelt:2006cw,Jaeckel:2006xm,Andriamonje:2007ew,Robilliard:2007bq,Zavattini:2007ee,Rubbia:2007hf}.

In the current paper we study pseudoscalar-photon mixing in the presence of correlated background magnetic fields in order to understand the results reported by Hutsemekers et al. \cite{Hutsemekers:1998,Hutsemekers:2000fv,Hutsemekers:2005iz} pertaining to the coherent alignment of quasar polarizations over Gpc scales. It has been shown earlier that propagation of radiation through the extragalactic medium, in the presence of a background magnetic field, can affect all Stokes parameters due to pseudoscalar-photon mixing. These changes have been investigated for radio \cite{Jain:1998kf,Jain:2002vx,Ralston:2003pf}, optical \cite{Hutsemekers:1998,Hutsemekers:2000fv,Jain:2002vx,Jain:2003sg,Hutsemekers:2005iz,Payez:2008pm,Piotrovich:2008iy}, and cosmic microwave background (CMB) \cite{Raffelt:1999tx,Mirizzi:2005ng,Raffelt:2006cw,Lee:2006za,Agarwal:2008ac} photons. We explicitly show that in our model it is possible to obtain nonzero correlations between the optical polarization of quasars separated by large distances in the sky. These nonzero correlations lead to an effect similar to the observed large-scale alignment of the polarizations.

We consider the extragalactic medium to be a large collection of magnetic domains, the background magnetic field in each domain being constant. This model was recently used in \cite{Agarwal:2008ac} where the extragalactic medium was considered as a large collection of uncorrelated magnetic domains in order to study the effects of pseudoscalar-photon mixing on CMB polarization. We extend the model of \cite{Agarwal:2008ac} to include correlations between magnetic fields in different domains, as motivated in \cite{Subramanian:2003sh,Seshadri:2005aa,Seshadri:2009sy}, and study the propagation of optical radiation from quasars through these domains.

The paper is organized as follows. In Sec. \ref{two} we outline the main concepts and equations of pseudoscalar-photon mixing and describe the domain propagation model. In Sec. \ref{three} we obtain magnetic field correlations between different domains. In Sec. \ref{four} we bring together results derived in the previous two sections and study alignments among the angles of polarization using a full numerical propagation model. We conclude with a brief discussion of our results and offer perspectives in Sec. \ref{five}. We also discuss an approximate analytical treatment to explain correlations in quasar polarizations in the Appendix.

\section{Pseudoscalar-photon mixing}
\label{two}

In this section, we consider the coupling of a light pseudoscalar to an electromagnetic field and discuss the propagation of electromagnetic waves in the presence of a background magnetic field. We first present a short review of the field of pseudoscalar-photon mixing in \ref{twoa} and subsequently describe our model of propagation in \ref{twob}.

\subsection{Basic concepts and equations}
\label{twoa}

The interaction Lagrangian for the coupling of pseudoscalars with an electromagnetic field can be written as,
\begin{eqnarray}
 \mathcal{L}_{\rm int} = \frac{g_\phi}{4} \phi F_{\mu\nu} \widetilde{F}^{\mu\nu},
\end{eqnarray}
where $g_\phi$ is the pseudoscalar-photon coupling constant, $\phi$ the pseudoscalar field, $F_{\mu\nu}$ the electromagnetic field tensor, and $\widetilde{F}_{\mu\nu} = \frac{1}{2} \epsilon_{\mu\nu\rho\sigma} F^{\rho\sigma}$ its dual. A single pseudoscalar is therefore effectively coupled to two photons. As shown in \cite{Das:2004qka,Das:2004ee}, the longitudinal component of the background magnetic field plays a negligible role in pseudoscalar-photon mixing. Since only photons polarized parallel to the transverse component of the background magnetic field ($\boldsymbol {\cal B}_{T}$) decay, this effect can lead to a change in polarization of the electromagnetic wave. Alternately, photons can mix with off-shell axions or axions may decay into photons, in either case leading to a change in polarization. Pseudoscalars can thus mix with photons from distant galaxies or the background CMB radiation, and lead to a rotation of polarization.

In order to describe the propagation of electromagnetic waves, we choose a coordinate system such that the $z$ axis lies along the direction of propagation, and the $x$ axis is parallel to the transverse component of the background magnetic field $\boldsymbol {\cal B}_{T}$. We define the gauge invariant quantity $\boldsymbol {A} = \boldsymbol {E}/\omega$ ($\boldsymbol {E}$ being the usual electric field and $\omega$ the frequency of radiation) and resolve it in components parallel ($A_{\parallel}$) and perpendicular ($A_{\perp}$) to the direction of $\boldsymbol {\cal B}_{T}$. This enables us to write the field equations solely in terms of $A_{\parallel}$ and $\phi$, since $A_{\perp}$ does not mix with the field $\phi$,
\begin{eqnarray}
    (\omega^2 + \partial_z^2) 
    \left(
    \begin{array}{c} 
    	A_\parallel(z) \cr \phi(z) 
    \end{array}
    \right)
    - M 
    \left( 
    \begin{array}{c}
    A_\parallel(z) \cr \phi(z) 
    \end{array}
    \right) = 0.
\label{eq:mixing}
\end{eqnarray}
The ``mass matrix'' or ``mixing matrix'', $M$, is given by,
\begin{eqnarray}
    M = 
    \left(
    \begin{array}{c c}
    	\omega_P^2 & - g_{\phi}{\cal B}_T\omega\cr
    	- g_{\phi}{\cal B}_T\omega & m_\phi^2
    \end{array}
    \right),
\label{eq:massmatrix}
\end{eqnarray}
where $\omega_P$ is the plasma frequency, $m_\phi$ the pseudoscalar mass, and ${\cal B}_T = |\boldsymbol {\cal B}_{T}|$.

The solution of the field equations follows from diagonalizing the above matrix equation using the mixing angle $\theta$ defined
by,
\begin{equation}
    \tan 2\theta = l g_{\phi}{\cal B}_T,
\end{equation}
where $l$ denotes the oscillation length,
\begin{equation}
    l = \frac{2\omega}{\omega_{P}^{2} - m_{\phi}^{2}}.
\label{eq:oscillationlen}
\end{equation}
In this paper we will be working in the limit of very small pseudoscalar mass, $m_\phi \ll \omega_P$, since for masses much heavier than this, the mixing with photons produces a negligible effect for intergalactic propagation, for the range of allowed parameters.

\subsection{Equations of polarization propagation}
\label{twob}

Polarization of radiation can be quantified in terms of Stokes parameters, which can be written as linear combinations of correlations between different components of the field $\boldsymbol {A}$. A convenient way to formulate the correlation functions between initial components of $\boldsymbol {A}$ and $\phi$ is to arrange them as elements of a physical density matrix $\rho(0)$, given by,
\begin{eqnarray}
    \rho(0) = 
    \left( 
    \begin{array}{c c c}
    	\langle A_{||}(0) A_{||}^{*}(0) \rangle & \langle A_{||}(0) A^{*}_{\bot}(0) \rangle & \langle A_{||}(0) \phi^{*}(0) \rangle
    	\cr \langle A_{\bot}(0) A^{*}_{||}(0) \rangle & \langle A_{\bot}(0) A^{*}_{\bot}(0) \rangle &  \langle  A_{\bot}(0) \phi^{*}(0) \rangle
    	\cr \langle \phi(0) A^{*}_{||}(0) \rangle & \langle  \phi(0) A^{*}_{\bot}(0) \rangle & \langle \phi(0) \phi^{*}(0) \rangle 
    \end{array}
    \right).
\label{eq:rho0}
\end{eqnarray}
The correlation functions propagated through a distance $z$ can then be expressed as,
\begin{equation}
    \rho(z) = P(z) \rho(0) P(z)^{-1},
\label{eq:rhoz}
\end{equation}
where the unitary matrix $P(z)$ describes the solution to the field equations, for a given mode $\omega$ \cite{Das:2004qka,Agarwal:2008ac}. In our coordinate system, $P(z)$ is given by,
\begin{eqnarray}
    P(z) = 
    e^{i(\omega+\Delta_{A})z} 
    \left( 
    \begin{array}{c c c}
    1-\gamma \sin^{2}\theta & 0 & \gamma \cos\theta \sin\theta
    \cr 0 & e^{-i\left[\omega+\Delta_{A}-(\omega^{2}-\omega_{P}^2)^{1/2}\right]z} & 0
    \cr \gamma \cos\theta \sin\theta & 0 & 1-\gamma\cos^{2}\theta
    \end{array} 
    \right),
\label{eq:pz}
\end{eqnarray}
where $\gamma = (1-e^{i\Delta z})$, $\Delta = \Delta_{\phi} - \Delta_{A}$, and $\Delta_{A}$, $\Delta_{\phi}$ are defined in terms of the frequency, $\omega$, and the eigenvalues, $\mu^2_{\pm}$, of the matrix $M$,
\begin{eqnarray}
    \Delta_{A} = \sqrt{\omega^2-\mu_{+}^2}-\omega, \\
    \Delta_{\phi} = \sqrt{\omega^2-\mu_{-}^2}-\omega.
\end{eqnarray}

For typical values of the electron density ($n_e \approx 10^{-8}$ cm$^{-3}$), and for optical radiation of quasars, with frequency $\nu \approx 10^{6}$ GHz (which corresponds to $l \approx 4$ Mpc, here $\nu = \omega/2\pi$), we have $\omega \gg \omega_P$, $m_{\phi}$, and $g_{\phi}{\cal B}_T$. Also, for typical values of the coupling, $g_{\phi} = 6\times10^{-11}$ GeV$^{-1}$, and magnetic field, $\mathcal{B}_{T} = 1$ nG, we can approximate $\Delta$ as,
\begin{equation}
    \Delta = \Delta_{\phi} - \Delta_{A} \approx \frac{1}{l} \sqrt{1 + \tan^{2}2\theta} = \frac{1}{l} \sec 2\theta.
\end{equation}

We now study the propagation of electromagnetic radiation through the intergalactic medium in the presence of pseudoscalar-photon mixing. Similar studies have been carried out earlier by several authors, e.g. \cite{Csaki:2001jk,Csaki:2001yk,Grossman:2002by,Mirizzi:2005ng,Mirizzi:2007hr,Agarwal:2008ac}. It is reasonable to model the medium as a large number of magnetic domains, with a uniform direction and strength of magnetic field in each domain. Additionally, the medium is assumed to have a uniform value of $\omega_{P}$. In the ensuing analysis we will use this model to explain correlations in the optical polarization of quasars as reported by Hutsemekers et al. \cite{Hutsemekers:1998,Hutsemekers:2000fv,Hutsemekers:2005iz}. A crucial component of our analysis will be the inclusion of correlations between magnetic fields in successive domains which has hitherto not been considered in any such study to the best of our knowledge. Details of these correlations will be presented in the next section.

The transverse magnetic field $\boldsymbol {\cal B}_{T}$ in the $i^{\rm th}$ domain is taken to be oriented at an angle $\beta_{i}$ with respect to the $x$ axis (the ``parallel axis'') of the external coordinate system. Starting with the density matrix $\rho(0) = {\rm diag}(1,0,0)$, corresponding to an initially unpolarized electromagnetic wave, the density matrix is propagated through each domain using (\ref{eq:rhoz}). After propagation through each domain, the electromagnetic wave vector is rotated back in order to account for the change in direction of the transverse magnetic field from one domain to another. The propagation through $n$ domains of size $z$ each (the $n^{\rm th}$ one being closest to the Earth) gives us an expression for $\rho_{n}(Z=nz)$ \cite{Agarwal:2008ac},
\begin{eqnarray}
	\rho_{n}(Z=nz) & = & R^{-1}(\beta_{n})P(z)R(\beta_{n})R^{-1}(\beta_{n-1})P(z)R(\beta_{n-1})R^{-1}(\beta_{n-2}) \ ... \ R^{-1}(\beta_{1})P(z)R(\beta_{1}) \cr  
	&  & \times \; \rho(0)R^{-1}(\beta_{1})P^{-1}(z)R(\beta_{1})R^{-1}(\beta_{2})P^{-1}(z)R(\beta_{2}) \ ... \ R^{-1}(\beta_{n})P^{-1}(z)R(\beta_{n}),
\label{eq:rhon}
\end{eqnarray}
where $R(\beta_{m})$ is the rotation matrix that acts only on the two-dimensional space transverse to the propagation direction; i.e., it represents a rotation by the angle $\beta_{m}$ about the $z$ axis. Explicit expressions for the propagation are given in the appendixes of \cite{Agarwal:2008ac}.

We also give here expressions for the reduced Stokes parameters in terms of different components of the density matrix, for use later in the paper,
\begin{subequations}
\bea
	I(z) & = & \rho_{11}(z) + \rho_{22}(z) = \langle A_{||}(z) A_{||}^{*}(z) \rangle + \langle A_{\bot}(z) A_{\bot}^{*}(z) \rangle, \\
	Q(z) & = & \rho_{11}(z)-\rho_{22}(z) = \langle A_{||}(z) A_{||}^{*}(z) \rangle - \langle A_{\bot}(z) A_{\bot}^{*}(z) \rangle, \\
	U(z) & = & \rho_{12}(z) + \rho_{21}(z) = \langle A_{||}(z) A_{\bot}^{*}(z) \rangle + \langle A_{\bot}(z) A_{||}^{*}(z) \rangle, \\
	V(z) & = & i(\rho_{12}(z) - \rho_{21}(z)) = i(\langle A_{||}(z) A_{\bot}^{*}(z) \rangle - \langle A_{\bot}(z) A_{||}^*(z) \rangle ).
\eea
\label{stokes}
\end{subequations}
Also, the linear polarization angle $\psi$ and the degree of polarization $p$ are given in terms of Stokes parameters by,
\bea
	\tan 2\psi & = & U/Q,
\label{psi} \\
	p & = & \sqrt{Q^2 + U^2 + V^2}/I.
\eea

\section{Magnetic field correlations}
\label{three}

We now calculate the correlations between components of the magnetic field in different domains. The analysis derives from the magnetic field spectrum $M(k)$ discussed in \cite{Subramanian:2003sh,Seshadri:2005aa,Seshadri:2009sy}, which can be defined using magnetic correlations of the form,
\bea
	\langle b_{i}({\boldsymbol k}) b^{*}_{j} ({\boldsymbol q}) \rangle & = & \delta_{{\boldsymbol{k,q}}} P_{ij}({\boldsymbol k}) M(k) \nonumber \\
	& = & \delta_{{\boldsymbol{k,q}}} \sigma^{2}_{ij}({\boldsymbol k}),
\label{eqn1}
\eea
where $\sigma^{2}_{ij}({\boldsymbol k}) = P_{ij}({\boldsymbol k}) M(k)$ and $k = |{\boldsymbol k}|$. Here ${\boldsymbol{b}(\boldsymbol {k})}$ is the Fourier transform of the present day magnetic field ${\boldsymbol{B}(\boldsymbol{r})}$ and $P_{ij}({\boldsymbol k}) = \left(\delta_{ij} -\frac{k_{i} k_{j}}{k^2}\right)$ is the projection operator included to be consistent with a divergenceless magnetic field. Although this distribution was originally proposed for primordial magnetic fields, it is justified to use it here since on galactic and larger scales the magnetic field simply redshifts away, as discussed in \cite{Seshadri:2005aa,Seshadri:2009sy,Jedamzik:1996wp,Subramanian:1997gi}. The magnetic field in real space is defined as,
\bea
	B_{j}({\boldsymbol r}) = \frac{1}{V} \sum b_{j}({\boldsymbol k}) e^{i {\boldsymbol k}.{\boldsymbol r}},
\label{IFT}
\eea
where $V$ is the volume in real space. In accordance with \cite{Subramanian:2003sh}, we consider a power-law dependence for
$M(k)$,
\be
    M(k) = A k^{n_{B}}\;;\;\;\;\;\; n_{B} > -3,
    \label{eqn2}
\ee
where $n_{B}$ is the power spectral index, and use a sharp $k$-space filter (\textit{window function}) of the form,
\begin{eqnarray}
    W(\theta) =
    \left\{
    \begin{array}{c c}
        1 \ \ & \theta < 1 \\
    0 \ \ & \theta > 1
  \end{array}
  \right..
\end{eqnarray} 

Using the above relations and taking the continuum limit by replacing $\sum_{\boldsymbol k}$ by $\frac{V}{(2\pi)^{3}} \int d^{3}{\boldsymbol k}$, we find that the spatial correlation between magnetic field components at two different points in space separated by a distance ${\boldsymbol{r}}'$ is given by,
\begin{eqnarray}
    \langle B_{i}({\boldsymbol{r}+\boldsymbol{r}'}) B_{j}({\boldsymbol r}) \rangle = \frac{1}{V} \int \frac{d^{3}{\boldsymbol k}}{(2\pi)^{3}} e^{i \boldsymbol{k}.{\boldsymbol{r}'}} \sigma^{2}_{ij}({\boldsymbol k}) W^{2}(kr_{G}), 
\label{spatcorr}
\end{eqnarray}
where $r_{G}$ is the ``galactic'' scale, taken to be 1 Mpc here. The constant $A$ in (\ref{eqn2}) is evaluated by assuming a smooth variation of the field ${\boldsymbol{B}(\boldsymbol{r})}$ over the scale $k_{G} = r_{G}^{-1} = 1 \; \textrm{Mpc}^{-1}$. Further, we consider a completely isotropic distribution of magnetic fields for simplicity, for which $P_{ij}({\boldsymbol k})$ = $(2/3)\delta_{ij}$. In view of the above assumptions, the spatial correlation calculated over a sphere of radius $r_{G} = 1$ Mpc gives the constant $A$ as,
\begin{eqnarray}
    A & = & V\pi^2 B_{0}^{2} \frac{(3 + n_{B})}{k_{G}^{3 + n_{B}}},
    \label{eqn3}
\end{eqnarray}
where $B_{0}$ is the fiducial constant magnetic field whose value is assumed to be 1 nG \cite{Seshadri:2009sy,Yamazaki:2010nf}. It may be appropriate to impose a large distance cutoff, $r_{\rm max}$, on the correlation for distances comparable to the size of the system (the Universe). In such a case the lower limit of the integral in (\ref{spatcorr}) will be $k_{\rm min} = r_{\rm max}^{-1}$; here we assume that this cutoff is small enough and can be approximated as zero.

We now use (\ref{spatcorr}) to obtain the correlations between magnetic field values in different domains. Consider the $m^{\rm th}$ and $n^{\rm th}$ domains along a given line of sight, separated by the distance $r_{mn}$. The transverse magnetic fields (of strength $\mathcal{B}_{T}$) in these domains are aligned at angles $\beta_{m}$ and $\beta_{n}$ with respect to the fixed external coordinate system, respectively. Hence the correlation between their components in the fixed coordinate system can be computed
as,
\bea
    \mathcal{B}_{T}^{2} \langle \cos\beta_{m} \cos\beta_{n} \rangle & = & \int_{0}^{k_{G}} \frac{dk}{k} \Delta_{b}^{2}(k) \left( \frac{\sin kr_{mn}}{kr_{mn}} \right), \ \ \ \ 
    \label{eqn4}
\eea
using $P_{ij}({\boldsymbol k}) = \delta_{ij}$ (since for transverse magnetic fields $k_{i} = k_{j} = 0$). Here $\Delta_{b}^{2}(k) = \frac{1}{V} \frac{k^{3}}{2 \pi^2 } M(k) = \frac{B_{0}^{2}}{2} (3 + n_{B}) \left(\frac{k}{k_{G}}\right)^{3 + n_{B}}$ is the power per logarithmic interval in $k$-space residing in the magnetic field. Therefore,
\begin{eqnarray}
    \mathcal{B}_{T}^{2} \langle \cos\beta_{m} \cos\beta_{n} \rangle & = & \frac{1}{2} B_{0}^2 \left(\frac{r_{G}}{r_{mn}}\right)^{3+n_{B}} \Bigg[ (3+n_{B}) \int_{0}^{r_{mn}/r_{G}} x^{1+n_{B}} \sin x \ dx \Bigg],
    \label{eqn5b}
\end{eqnarray}
where the term in square brackets is of order unity. For example, for $n_{B} = -2.37$, which was reported as the best fit value in the statistical and numerical analysis of the matter and CMB power spectrum in \cite{Yamazaki:2010nf}, an explicit numerical evaluation yields a value for this term between 0.96 and 1.33 for $r_{mn}/r_{G} = 1$ and large $r_{mn}/r_{G}$, respectively. Further, for $n_{B} \rightarrow -3.0$ this term rapidly approaches unity.

For the completely isotropic case considered in this analysis we have $P_{ij}({\boldsymbol k}) \propto \delta_{ij}$, leading to a cancellation of any cross correlations between transverse components of the magnetic field, i.e., $\langle \cos\beta_{m} \sin\beta_{n} \rangle = 0$. Also, the invariance of correlations under rotation of the external coordinate system immediately implies an equality between $\langle \cos\beta_{m} \cos\beta_{n} \rangle$ and $\langle \sin\beta_{m} \sin\beta_{n} \rangle$. The final set of correlations can therefore be summarized as:
\begin{subequations}
\begin{align}
   \langle \cos\beta_{m} \cos\beta_{n} \rangle & = \frac{1}{2} \frac{B_{0}^2}{\mathcal{B}_{T}^{2}} \left(\frac{r_{G}}{r_{mn}}\right)^{3+n_{B}}, \\
   \langle \cos\beta_{m} \sin\beta_{n} \rangle & = 0, \\
   \langle \sin\beta_{m} \sin\beta_{n} \rangle & = \frac{1}{2} \frac{B_{0}^2}{\mathcal{B}_{T}^{2}} \left(\frac{r_{G}}{r_{mn}}\right)^{3+n_{B}}.
\end{align}
\label{corr2point}
\end{subequations}


\section{Correlations in the optical polarization of quasars}
\label{four}

In this section we use the domain propagation model described in Sec. \ref{two}, with magnetic fields correlated according to the correlation functions discussed in Sec. \ref{three}, to understand the large-scale alignment of quasar polarizations observed by Hutsemekers et al. \cite{Hutsemekers:1998,Hutsemekers:2000fv,Hutsemekers:2005iz}. We perform a detailed numerical analysis of the correlations by direct integration of the propagation equations, using a magnetic field distribution that follows from our discussion in the previous section. We also refer interested readers to the Appendix, where we have included an analytical approach to estimate such correlations in quasar polarizations in the regime $z/l \ll 1$ and $\theta \ll 1$.

We first extract a one-dimensional (1D) magnetic field distribution along a given line of sight from (\ref{spatcorr}), by integrating over the transverse components of ${\boldsymbol k}$. This gives
\bea
	\langle B_{i}(z+z') B_{j}(z) \rangle = \frac{1}{V} \int_{0}^{k_{G}} \frac{dk_{z}}{2\pi} e^{i k_{z}z'} \sigma^{2}_{ij}(k_{z}),
\eea
where $\sigma^{2}_{ij}(k_{z}) = \delta_{ij}\widehat{M}(k_{z})$. The 1D magnetic field spectrum $\widehat{M}(k_{z})$ is now given by,
\bea
	\widehat{M}(k_{z}) = \frac{A}{3\pi} \left( \frac{k_{G}^{2+n_{B}} - k_{z}^{2+n_{B}}}{2+n_{B}} \right).
\eea
A distribution function compatible with the above correlations can be written as,
\begin{eqnarray}
	f(b_{i}(k_{z}),b_{j}(k_{z})) = N_{i}(k_{z}) N_{j}(k_{z}) \ {\rm exp} \left[-\left(\frac{b_{i}^{2}(k_{z}) + b_{j}^{2}(k_{z})}{2\widehat{M}(k_{z})}\right) \right],
\end{eqnarray}
where $N_{i}(k_{z})$, $N_{j}(k_{z})$ are the normalizations. This represents an uncorrelated Gaussian distribution for two components of the magnetic field in Fourier space, corresponding to the wave vector $k_{z}$.

Using the above distribution, we generate the transverse Fourier components $b_{x}(k_{z})$ and $b_{y}(k_{z})$ of the magnetic field for the discretized wave vector $k_{z}$. This can be done for each domain in Fourier space independently as the distribution is uncorrelated for different $k_{z}$. The magnetic field components $B_{x}(z)$ and $B_{y}(z)$ in real space can then be obtained by implementing the inverse Fourier transform in (\ref{IFT}). The transverse component of the magnetic field is characterized by the magnitude $\mathcal{B}_{T} = \sqrt{B_{x}^{2}(z) + B_{y}^{2}(z)}$ and the angle $\beta = \tan^{-1}(B_{y}(z)/B_{x}(z))$ that it makes with the $x$ axis of the external coordinate system. We repeat this exercise for every domain to obtain the complete magnetic field landscape along a given line of sight.

\begin{figure}[!t]
  \centering{
    \includegraphics[width=4.5in,angle=0]{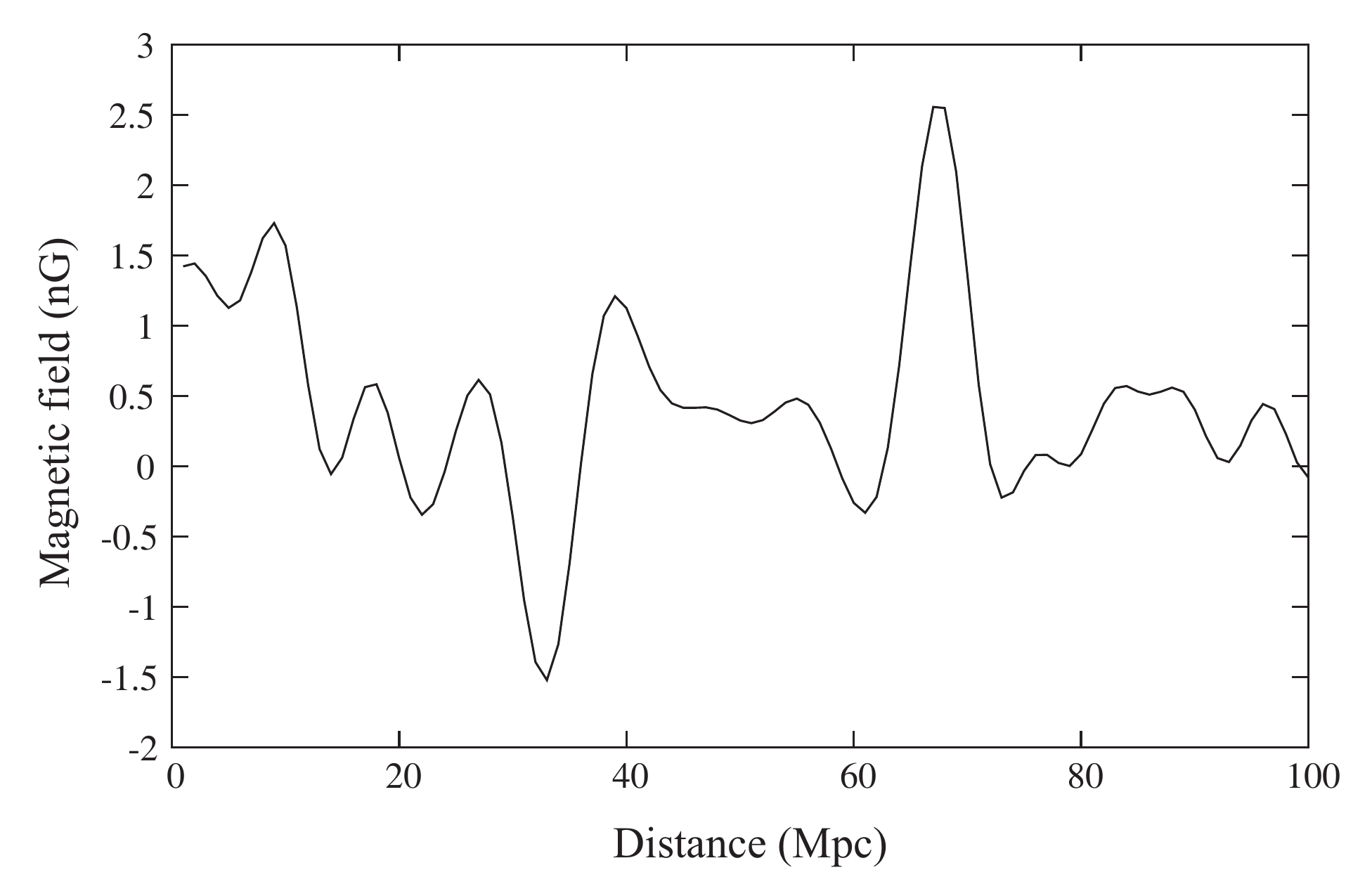}
    \caption{A random realization of one of the components of the correlated magnetic field as a function of distance.} 
   \label{fig1}}
\end{figure}

We point out that the magnetic field changes relatively slowly over small distances of order 10 Mpc. This can be seen in Fig. {\ref{fig1}} where we plot a random realization of one of the components of the correlated magnetic field. Hence our line of sight calculation applies approximately even if quasars lie at slightly different angles. For larger angular separations we require a full 3D simulation which is very time-consuming and we postpone this for future research.

Next, using the above model, we study propagation along a given line of sight through $2^{12}$ domains of size $z = 1$ Mpc each, with 400 equally spaced quasars - the nearest one being 10 Mpc away and the farthest one being 4 Gpc away from the Earth. The initial electromagnetic radiation of each quasar is assumed to be unpolarized and the parameters $g_{\phi}$ and $\nu$ are chosen to have typical values $6 \times 10^{-11}$ GeV$^{-1}$ and $10^{6}$ GHz, respectively. The simplest statistic to test whether our model predicts an alignment of quasar polarization vectors is to calculate the dispersion of the polarization vector of a quasar with respect to its nearest neighbors \cite{Jain:2003sg}. For this exercise, we group quasars into $n_{s}$ sets of $n_{v}$ quasars each, and calculate the mean value of the vector $[\cos(2\psi_{i}),\sin(2\psi_{i})]$, $\psi_{i}$ being the polarization angle [calculated using (\ref{psi})], for each set. A measure of the dispersion $d_{k}$ of vectors in each set is given by,
\bea
	d_{k} = \frac{1}{n_{v}} \sum_{i=1}^{n_{v}} \cos[2(\psi_{i} - \psi_{k})],
\eea
where $\psi_{k}$ is the mean polarization angle in the $k^{\rm th}$ set. The statistic defined as,
\bea
	S^{P}_{D} = \frac{1}{n_{s}} \sum_{k=1}^{n_{s}} d_{k},
\label{spd}
\eea
gives us a measure of the dispersion in the data. A large value of $S^{P}_{D}$ indicates a strong alignment between polarization vectors.

\begin{figure}[!t]
  \centering{
    \includegraphics[width=4.5in,angle=0]{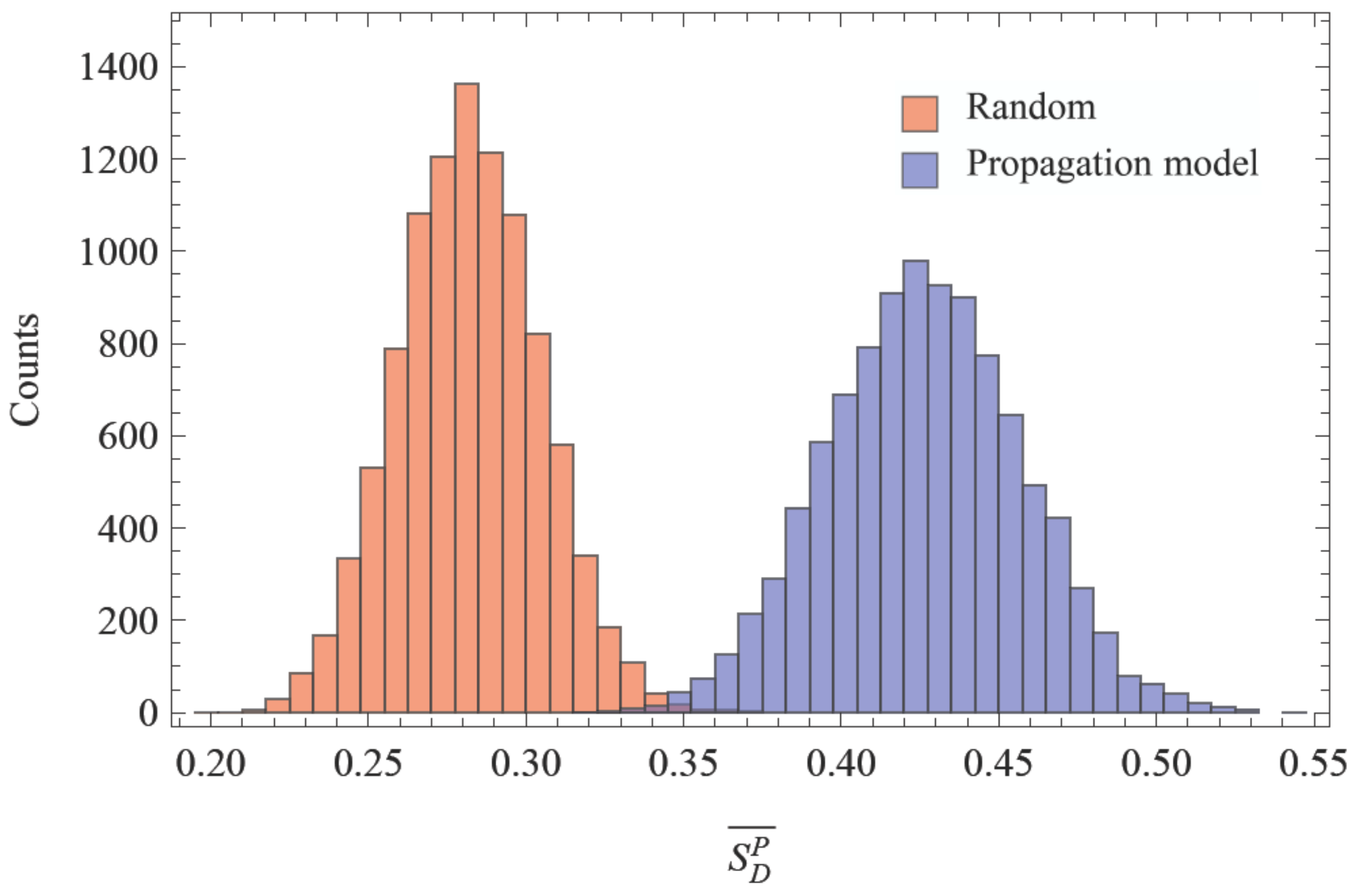}
    \caption{Histograms of the statistic $S^{P}_{D}$ with the number of nearest neighbors $n_{v} = 10$, for $10^{4}$ random samples (left-red) and $10^{4}$ realizations of the magnetic field distribution in the propagation model proposed in the paper (right-blue), using parameter values $n_{e} = 10^{-8}$ cm$^{-3}$, $n_{B} = -2.37$, $\nu = 10^{6}$ GHz, and $g_{\phi} = 6 \times 10^{-11}$ GeV$^{-1}$. The larger value of the mean $\overline{S^{P}_{D}}$ obtained for the propagation model, compared to the random sample, indicates preferential alignment of quasar polarization vectors, as reported by Hutsemekers et al. \cite{Hutsemekers:1998,Hutsemekers:2000fv,Hutsemekers:2005iz}.} 
   \label{fig2}}
\end{figure}

\begin{figure}[!t]
  \centering{
    \includegraphics[width=3.25in,angle=0]{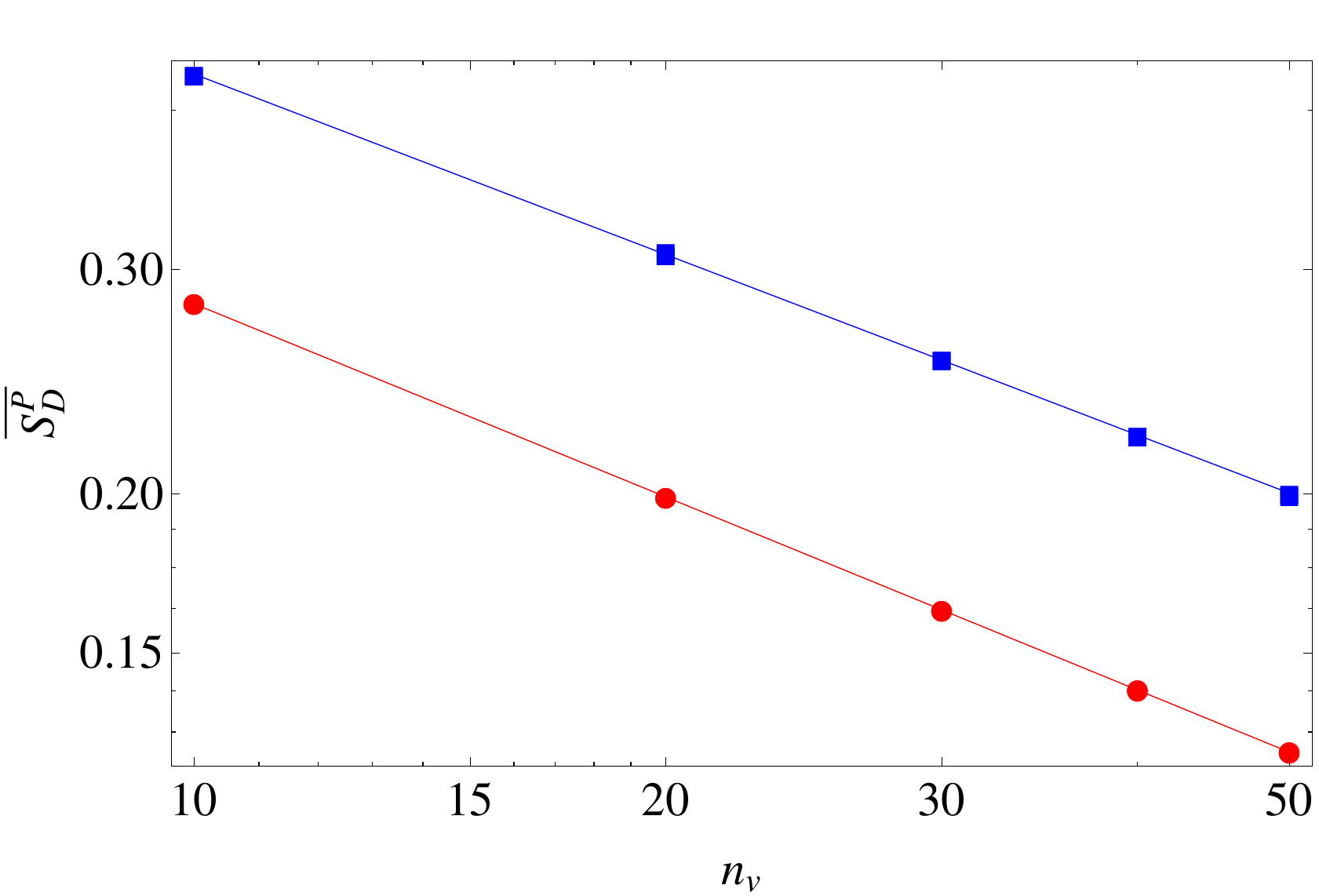}
    \includegraphics[width=3.25in,angle=0]{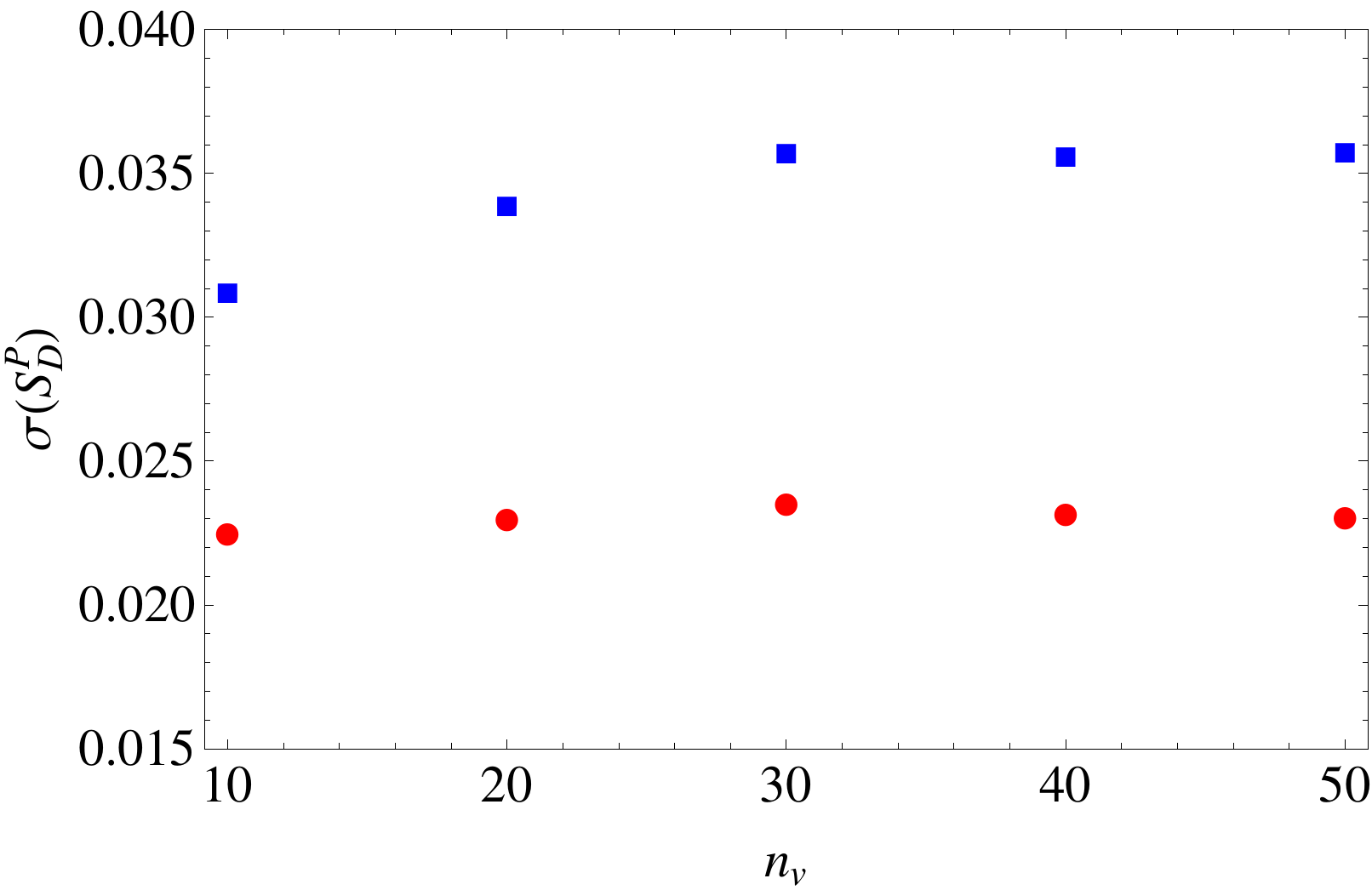}
    \caption{The mean (left panel) and standard deviation (right panel) of the statistic $S^{P}_{D}$ as a function of nearest neighbors $n_{v}$ for $10^{4}$ random samples (red circles) and $10^{4}$ realizations of the magnetic field distribution in our model of propagation (blue squares), using identical parameter values as Fig. \ref{fig2}. The variation of the mean $\overline{S^{P}_{D}}$ with $n_{v}$ has been plotted on a log-log scale. The lines are fits generated using a power-law dependence of the form $ax^{b}$ [where $b \approx -0.50$, $a \approx 0.90$ for the random sample (red); $b \approx -0.47$, $a \approx 1.26$ for the propagation model using correlated magnetic fields (blue)].} 
   \label{fig3}}
\end{figure}

In Fig. \ref{fig2} we show results of the $S^{P}_{D}$ statistic with $n_{v} = 10$, obtained using polarization angles from our propagation model and compare it with data generated using a random distribution of the polarization angles. Also, we have checked that the distribution of polarization angles obtained using random magnetic fields, when quasars along a given line of sight are at slightly different angles so that each experiences completely random magnetic fields on propagation, mimics a random distribution of polarization angles. Since the mean value $\overline{S^{P}_{D}}$ of the statistic in (\ref{spd}) is clearly larger for data obtained using the propagation model, this provides conclusive evidence of a preferred alignment among quasar polarization vectors. Further, we expect the mean value of this statistic for a random sample to be proportional to $1/\sqrt{n_{v}}$ \cite{Jain:2003sg}. We indeed find that results for the random sample are well described by the fit $\overline{S^{P}_{D}} \approx 0.90/\sqrt{n_{v}}$ (Fig. \ref{fig3}). Further, for the propagation model, $\overline{S^{P}_{D}} \approx 1.26/n_{v}^{0.47}$. This approximately $1/\sqrt{n_{v}}$ behavior in our model suggests that the angles of polarization are preferentially aligned in each neighborhood, yet span the whole range of possible angles as in the case of a random distribution of angles.

\begin{figure}[!t]
  \centering{
    \includegraphics[width=4.5in,angle=0]{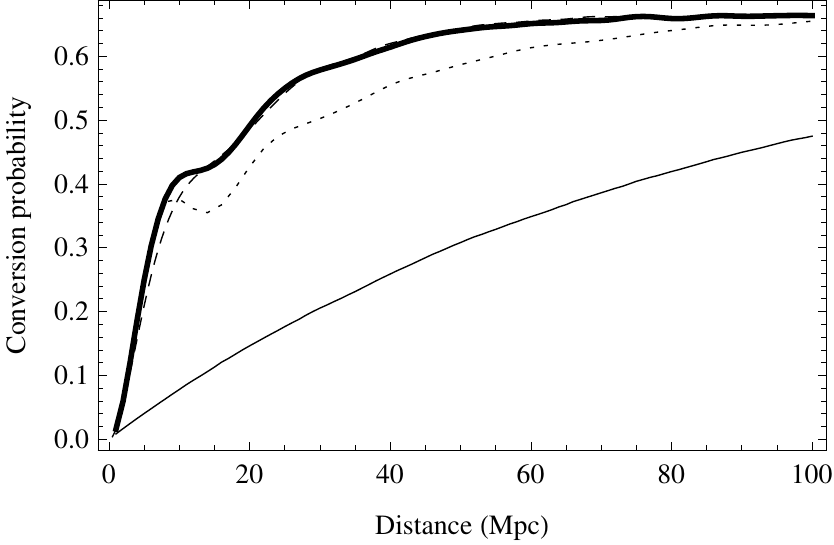}
    \caption{Axion conversion probability as a function of distance traveled, for optical radiation from a single initially unpolarized quasar 100 Mpc away from us, as it encounters correlated (solid thick black curve for $k_{G} = 1$ Mpc$^{-1}$, small-dashed black curve for $k_{G} = 0.5$ Mpc$^{-1}$, large-dashed black curve for $k_{G} = 2$ Mpc$^{-1}$) or random (solid thin black curve) magnetic field distributions on its journey. Each curve is generated with parameter values $n_{e} = 10^{-8}$ cm$^{-3}$, $n_{B} = -2.37$, $\nu = 10^{6}$ GHz, $g_{\phi} = 6 \times 10^{-11}$ GeV$^{-1}$, and is averaged over $10^{4}$ realizations of the respective magnetic field distribution.}
   \label{fig4}}
\end{figure}

A related effect of correlated magnetic fields in our propagation model is that they lead to a larger axion conversion probability, defined as
\bea
	P_{\gamma \rightarrow \phi} = \langle \phi(z)\phi^{*}(z) \rangle,
\eea
compared to that obtained using a model with random magnetic fields. Therefore it is more likely for photons to decay into axions, over a given distance, in the presence of a correlated magnetic field distribution. This can be seen in Fig. \ref{fig4} where we show the development of the conversion probability with distance for an initially unpolarized quasar, 100 Mpc away from us, as optical radiation travels through correlated and random magnetic field distributions.

We finally discuss how our results change with a different choice of $k_G$ or equivalently $r_G$. Here we assume that the magnetic field is uniform in each domain, and the size of each domain is equal to $r_G$. In the results presented so far we have chosen $r_G=1$ Mpc. A larger value of $r_G$ will induce correlations over larger distances, hence we would naively expect a stronger signal for larger $r_G$ or equivalently smaller $k_G$. However, the phenomenon is nonlinear and complicated. For example, the photon to pseudoscalar conversion probability does not increase monotonically but oscillates. This is because of the reverse process of pseudoscalar to photon conversion, once the population of pseudoscalars becomes sufficiently large. Furthermore the orientation of linear polarization does not evolve proportionately to the conversion probability. Hence it is not possible to anticipate how the results depend on $k_G$. In Fig. \ref{fig4} we also show the conversion probability for $k_G=2$ Mpc$^{-1}$ and $k_G=0.5$ Mpc$^{-1}$. We see that the conversion probability slightly decreases in both cases. On calculating the $S^{P}_{D}$ statistic for these cases we find that for $n_{v} = 10$ nearest neighbors, $\bar{S}^{P}_{D} \approx 0.43$ for $k_{G} = 1$ Mpc$^{-1}$, $0.44$ for $k_{G} = 2$ Mpc$^{-1}$, and $0.48$ for $k_{G} = 0.5$ Mpc$^{-1}$. Therefore for larger $r_{G}$ there are higher correlations although the conversion probability is smaller due to the conversion of pseudoscalars into photons. For smaller $r_{G}$, on the other hand, the correlation is only feebly affected, and the conversion probability decreases for most values of the distance, as expected.

\section{Discussion}
\label{five}

A very striking alignment in the optical polarization of quasars has been observed over cosmologically large distances ($\sim$ 1 Gpc) by Hutsemekers et al. \cite{Hutsemekers:1998,Hutsemekers:2000fv,Hutsemekers:2005iz} at both low ($z \sim 0.5$) and high redshifts ($z \sim 1.5$). It is unlikely that quasar polarizations are intrinsically aligned with one another over such large distances since this appears to be in conflict with the basic assumptions of isotropy and homogeneity of the Universe. It has been proposed in the past that these observations may be explained in terms of a propagation effect related to axion-photon mixing. In the current paper we have explored this possibility in detail, and studied the propagation of electromagnetic radiation through a large number of magnetic domains. 

We have shown that correlations between components of the background magnetic field in various domains induce correlations between the linear polarization of different quasars. Our analysis explicitly shows that quasar polarizations are expected to be aligned in such a scenario due to the presence of significant correlations between them, as observed by Hutsemekers et al. \cite{Hutsemekers:1998,Hutsemekers:2000fv,Hutsemekers:2005iz}. A calculation of the dispersion of polarization with respect to nearest neighbors in a model of propagation along a given line of sight shows significant alignments compared to a random sample. A direct comparison with observations and further study of the effect clearly requires numerical simulations of the complete 3D propagation model that we have proposed. We leave such an analysis to future work.

We further note that a calculation of correlations in the circular polarization, Stokes $V$ parameter, is also expected to yield a significant result. A detailed calculation of this parameter is postponed to future research. This may prove to be an important signature of pseudoscalar-photon mixing in the observed alignment of quasar polarizations, as noted in \cite{Hutsemekers:2008iv}.


\section*{Acknowledgements}

The authors wish to thank Douglas W. McKay, Moninder S. Modgil, John P. Ralston, and T. R. Seshadri for providing valuable insights. N.A. is also grateful to Grant J. Mathews for very useful discussions on magnetic field correlations and for sharing unpublished results on this.


\appendix
\section*{Appendix: Analytical treatment of the correlations}
\label{appendix}

\renewcommand{\theequation}{A\arabic{equation}}
\setcounter{equation}{0}

In this appendix we discuss an analytical approach to estimate the correlations in quasar polarizations for small domain size ($z/l \ll 1$) and small mixing angle ($\theta \ll 1$). In this limit, we expand the 22 element of (\ref{eq:pz}) to leading order,
\begin{equation}
    e^{-i \left[\omega+\Delta_{A}-(\omega^{2}-\omega_{P}^2)^{1/2}\right] z} \approx 1 - i \frac{z}{2l} (1-\sec2\theta),
\label{eq:22}
\end{equation}
where we have omitted the overall phase factor. We can further write the propagation matrix as the sum of a quasiunit matrix and a matrix proportional to $\sin\theta$,
\begin{equation}
	P(z) = \mathcal{I}(z) + \sin\theta \; \mathcal{P}(z,\theta).
\label{eq:defP}
\end{equation}
Suppressing the irrelevant overall phase in (\ref{eq:pz}), the matrices $\mathcal{I}(z)$ and $\mathcal{P}(z,\theta)$ are given by,
\begin{equation}
	\mathcal{I}(z) =  {\rm diag} (1,1,e^{i\Delta z}),
\label{eq:unit}
\end{equation}
and,
\begin{equation}
	\mathcal{P}(z,\theta) = 
	\left(
	\begin{array}{c c c}
	-\gamma \sin\theta & 0 & \gamma \cos\theta
	\cr 0 & i\frac{z}{l}\sin\theta\sec2\theta & 0 
	\cr \gamma \cos\theta  & 0 & \gamma \sin\theta
	\end{array}
	\right).
\label{eq:calP}
\end{equation}
We can now express $\rho_{n}(Z=nz)$ as a power series in $\sin\theta$. For an initially unpolarized electromagnetic wave corresponding to a density matrix $\rho(0) = \textrm{diag}(1,1,0)$, the calculation outlined above gives the following leading order terms in different powers of $\sin\theta \; \mathcal{P}(z,\theta)$,
\begin{eqnarray}
	\rho_{n}^{[0]}(Z) & = & \rho(0),
\label{eq:leading0} \\
	\rho_{n}^{[1]}(Z) & = & \sum_{j=1}^{n} \rho_{n}^{[1]}(Z,j),
\label{eq:leading1} \\
  \rho_{n}^{[2]}(Z) & = & \sum_{k=2}^{n} \sum_{j=1}^{k-1} \rho_{n}^{[2]}(Z,j,k),
\label{eq:leading2}
\end{eqnarray}
where,
\begin{eqnarray}
    \rho_{n}^{[1]}(Z,j) =
    \left(
    \begin{array}{c c c}
    -2\sin^2\theta \cos^2\beta_j(1-\cos\Delta z) & -\sin^2\theta \sin2\beta_j (1-\cos\Delta z)& \gamma^{*} e^{-i\Delta z(n-j)} \cos\beta_j \sin(2\theta)/2    \cr
    -\sin^2\theta \sin2\beta_j(1-\cos\Delta z) & -2\sin^2\theta \sin^2\beta_j(1-\cos\Delta z) & \gamma^{*} e^{-i\Delta z(n-j)}\sin\beta_j \sin(2\theta)/2   \cr
    \gamma e^{i\Delta z(n-j)} \cos\beta_j \sin(2\theta)/2 & \gamma e^{i\Delta z(n-j)}\sin\beta_j \sin(2\theta)/2 & 0
    \end{array}
    \right),
\label{eq:rho1}
\end{eqnarray}
and,
\begin{subequations}
\bea
    \rho_{n,11}^{[2]}(Z,j,k) & = &
-4 \sin^2\theta\cos^{2}\theta(1-\cos\Delta z)\cos[\Delta z(k-j)] \cos\beta_k \cos\beta_j, \ \ \\
    \rho_{n,12}^{[2]}(Z,j,k) & = &
-2 \sin^2\theta\cos^{2}\theta(1-\cos\Delta z)\cos[\Delta z(k-j)] \Big( \sin(\beta_k+\beta_j)-
i\sin(\beta_k-\beta_j)\tan[\Delta z(k-j)] \Big), \ \ \\
    \rho_{n,21}^{[2]}(Z,j,k) & = &
-2 \sin^2\theta\cos^{2}\theta(1-\cos\Delta z)\cos[\Delta z(k-j)] \Big( \sin(\beta_k+\beta_j)+
i\sin(\beta_k-\beta_j)\tan[\Delta z(k-j)] \Big), \ \ \\
    \rho_{n,22}^{[2]}(Z,j,k) & = &
-4 \sin^2\theta\cos^{2}\theta(1-\cos\Delta z)\cos[\Delta z(k-j)] \sin\beta_k \sin\beta_j . \ \ 
\eea
\label{eq:nextleadingjk}
\end{subequations}
Here we only report elements of the $2 \times 2$ submatrix relevant to polarization for $\rho_{n}^{[2]}(Z,j,k)$. Further, we ignore higher-order terms in writing (\ref{eq:nextleadingjk}). This approximation is valid as long as $\sin^{2}\theta$ and $(z/l)\sin^{2}\theta$ are much smaller than unity.

For two quasars (along a given line of sight) at distances $Z_{1} = nz$ and $Z_{2} = mz$, where $n$ and $m$ are the number of domains of size $z$ that lie between us and the two quasars, we wish to estimate the correlation $\langle U(Z_{1}=nz)U(Z_{2}=mz) \rangle$. Using (\ref{stokes}) we find that,
\begin{eqnarray}
	\langle U(Z_{1})U(Z_{2}) \rangle & = & \langle \rho_{n,12}(Z_{1}) \rho_{m,12}(Z_{2}) \rangle + \ \langle \rho_{n,12}(Z_{1}) \rho_{m,21}(Z_{2}) \rangle \nonumber \\
    & & \ + \ \langle \rho_{n,21}(Z_{1}) \rho_{m,12}(Z_{2}) \rangle + \ \langle \rho_{n,21}(Z_{1}) \rho_{m,21}(Z_{2}) \rangle. \ \ \ \ \  
\label{corrU}
\end{eqnarray}
On removing the constant zeroth order part of the propagated density matrix we can write,
\begin{eqnarray}
    \rho_{n}(Z) & \approx & \rho^{[1]}_{n}(Z) + \rho^{[2]}_{n}(Z).
\end{eqnarray}
Using (\ref{eq:rho1}) and (\ref{eq:nextleadingjk}) in this expression to calculate (\ref{corrU}), we end up with four-point correlation functions in the sines and cosines of rotation angles. We assume that the magnetic field fluctuations at any point are Gaussian, so that we can express these four-point correlations in terms of two-point correlations using the following decomposition,
\begin{eqnarray}
	\langle S(r_{1})S(r_{2})S(r_{3})S(r_{4}) \rangle & = & \langle S(r_{1})S(r_{2}) \rangle \langle S(r_{3})S(r_{4}) \rangle + \langle S(r_{1})S(r_{3}) \rangle \langle S(r_{2})S(r_{4}) \rangle + \langle S(r_{1})S(r_{4}) \rangle \langle S(r_{2})S(r_{3}) \rangle, \ \ \ \ \ 
	\label{wick}
\end{eqnarray}
where $S(r)$ is some function (such as a Gaussian) that falls off sufficiently fast with $r$. Now we can further use (\ref{corr2point}), and write the correlation in (\ref{corrU}) as,
\begin{eqnarray}
    & & \left\langle U(Z_{1})U(Z_{2}) \right\rangle = 4\left( \frac{B_{0}}{\mathcal{B}_{T}} \right)^{4} \sin^{4}\theta(1-\cos\Delta z)^{2} \times \Bigg[ \sum_{i=1}^{n} \sum_{p=1}^{m} \left( \frac{r_{G}}{r_{ip}} \right)^{2(3+n_{B})} \nonumber \\  
    & & \ \ + \ 2\cos^{2}\theta \left\{ \sum_{i=1}^{n}\sum_{q=2}^{m}\sum_{p=1}^{q-1} \left( \frac{r_{G}^{2}}{r_{qi}r_{ip}} \right)^{3+n_{B}} \cos[\Delta z(q-p)] + \sum_{p=1}^{m}\sum_{j=2}^{n}\sum_{i=1}^{j-1} \left( \frac{r_{G}^{2}}{r_{jp}r_{pi}} \right)^{3+n_{B}} \cos[\Delta z(j-i)] \right\} \nonumber \\ 
    & & \ \ + \ \cos^{4}\theta \sum_{j=2}^{n}\sum_{i=1}^{j-1}\sum_{q=2}^{m}\sum_{p=1}^{q-1} \left\{ \left( \frac{r_{G}^{2}}{r_{qj}r_{pi}} \right)^{3+n_{B}} + \left( \frac{r_{G}^{2}}{r_{qi}r_{pj}} \right)^{3+n_{B}} \right\} \Big( \cos[\Delta z(j+q-i-p)] + \cos[\Delta z(j-q-i+p)] \Big) \Bigg]. \nonumber \\
\label{corrUfinal}
\end{eqnarray}
The above equation suggests that the linear polarization of quasars at different redshifts along a given line of sight is correlated, as observed by Hutsemekers et al. \cite{Hutsemekers:1998,Hutsemekers:2000fv,Hutsemekers:2005iz}. Here in using the correlations (\ref{corr2point}) we have set the term in square brackets in (\ref{eqn5b}) to unity. Further we choose a constant value of 1 nG for $\mathcal{B}_{T}$. These approximations affect the analytical result to within a factor of order unity.

The analytical calculation presented here makes several approximations and hence only provides a rough estimate of the correlations. Nonetheless, it is very important since it establishes the presence of correlations among polarizations of electromagnetic radiation from quasars separated by large distances. These correlations arise due to the presence of correlations in the intergalactic magnetic field. The analytical result also provides useful insight into how the correlations among polarizations depend on the intergalactic magnetic field. In order to study correlations beyond the regime of validity of the analytical method discussed here, a complete numerical treatment of propagation, as done in Sec. \ref{four} of this paper, is necessary.

\begin{figure}[!t]
  \centering{
    \includegraphics[width=4.5in,angle=0]{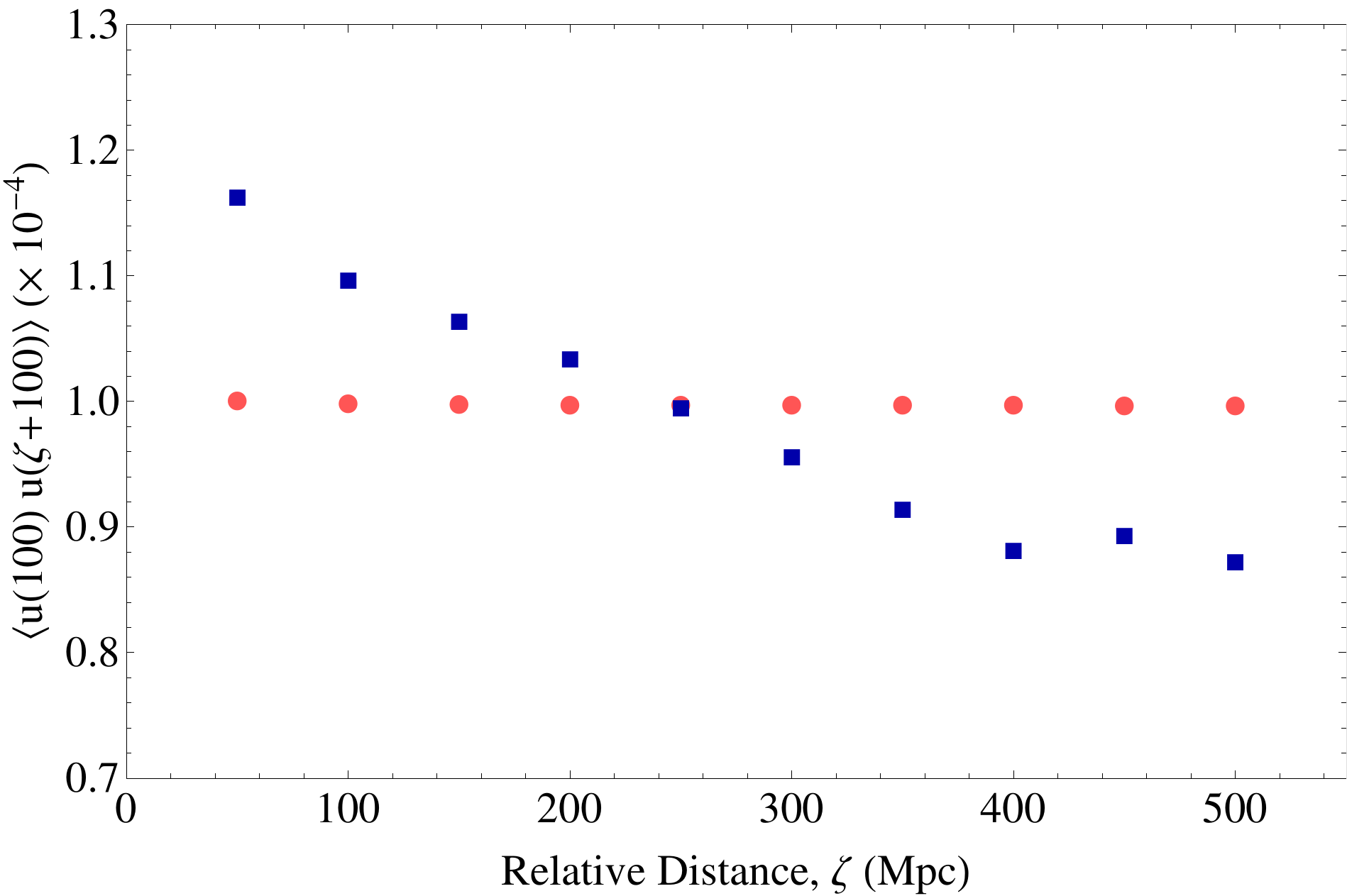}
    \caption{Comparison of analytical and numerical results for parameter values $n_{e} = 10^{-8}$ cm$^{-3}$, $n_{B} = -2.37$, $\nu = 10^{6}$ GHz, and $g_{\phi} = 6 \times 10^{-12}$ GeV$^{-1}$. The plot shows the correlation $\langle u(Z_{1} = 100 {\rm \ Mpc})u(Z_{2} = (100 + \zeta) {\rm \ Mpc}) \rangle$ for different values of the relative distance $\zeta$ between two quasars, where $u$ is the normalized value of the Stokes parameter $U$ with respect to the initial intensity, $I(0) = 2$. Red circles show values obtained using the analytical result (\ref{corrUfinal}) and blue squares are values obtained after the complete numerical propagation, averaged over $10^{4}$ realizations of the magnetic field distribution.} 
   \label{fig5}}
\end{figure}

It is also interesting to use the numerical propagation model in the regime where the analytical calculation holds, and compare the result with that obtained using (\ref{corrUfinal}). For this, we study propagation along a given line of sight through a total of $2^{10}$ domains of size $z = 1$ Mpc each, with 100 equally spaced quasars - the nearest one being 10 Mpc away and the farthest one being 1 Gpc away from the Earth. The initial electromagnetic radiation of each quasar is assumed to be unpolarized. We choose the following parameter values: electron density $n_{e} = 10^{-8}$ cm$^{-3}$, spectral index $n_{B} = -2.37$, frequency of radiation $\nu = 10^{6}$ GHz, coupling $g_{\phi} = 6 \times 10^{-12}$ GeV$^{-1}$, and $B_{0} = 1$ nG. This choice of parameters ensures that $z/l < 1$ and $\theta \ll 1$. Further, in the analytical result we set $\mathcal{B}_{T} = 1$ nG and calculate $r_{ip}$ as $r_{ip} = z|i-p|$, while $r_{ii} = r_{G}$. In the numerical propagation, the value of $\mathcal{B}_{T}$ and the angle $\beta$ in each domain are obtained from the magnetic field distribution discussed in Sec. \ref{four}. For the purpose of comparison with the analytical result, however, we set the magnitude of the magnetic field at every point to be equal to 1 nG in the numerical calculation as well. Direct calculation shows that allowing the magnitude to vary makes a significant difference in the final answer. Hence it does not provide a fair comparison with the analytical result where we have fixed the magnitude of the field in order to make the calculation tractable. Fig. \ref{fig5} shows a plot of analytical and numerical values of the correlation between the Stokes $U$ parameters of two quasars separated by a relative distance $\zeta$. 

We find that the analytical and numerical results agree to within about 20\%. The discrepancy between the two results may be attributed to several other approximations made in the analytical treatment, where (i) we have set the term in square brackets in (\ref{eqn5b}) to unity and (ii) we have ignored higher-order corrections. As a rough estimate, including the term in square brackets in (\ref{eqn5b}) increases the analytical values of correlations by $60\%-70\%$. In addition, we also observe that the numerical propagation model has a small dependence on the ``system size'' (i.e., the total number of domains chosen), unlike the analytical result in (\ref{corrUfinal}). This can be ascribed to dependence on the number of domains, of the discrete values of the wave vector $k_{z}$ used in the inverse Fourier transform. As the magnetic field values depend on $k_{z}$, this introduces a small system size dependence in the results. Hence we should expect the two results to agree only within a factor of about 2. We also point out that we expect best agreement for small distances of propagation. This is because for larger distances the higher-order terms, dropped in the analytical calculation, cannot be neglected. In this case we will have to sum over a large number of domains and the higher-order terms will become increasingly important. The approximation used in (\ref{eq:22}) will therefore break down.


\bibliographystyle{apsrev}
\bibliography{paper_prd}

\begin{thebibliography}{78}
\expandafter\ifx\csname natexlab\endcsname\relax\def\natexlab#1{#1}\fi
\expandafter\ifx\csname bibnamefont\endcsname\relax
  \def\bibnamefont#1{#1}\fi
\expandafter\ifx\csname bibfnamefont\endcsname\relax
  \def\bibfnamefont#1{#1}\fi
\expandafter\ifx\csname citenamefont\endcsname\relax
  \def\citenamefont#1{#1}\fi
\expandafter\ifx\csname url\endcsname\relax
  \def\url#1{\texttt{#1}}\fi
\expandafter\ifx\csname urlprefix\endcsname\relax\def\urlprefix{URL }\fi
\providecommand{\bibinfo}[2]{#2}
\providecommand{\eprint}[2][]{\url{#2}}

\bibitem[{\citenamefont{Peccei and Quinn}(1977{\natexlab{a}})}]{Peccei:1977hh}
\bibinfo{author}{\bibfnamefont{R.~D.} \bibnamefont{Peccei}} \bibnamefont{and}
  \bibinfo{author}{\bibfnamefont{H.~R.} \bibnamefont{Quinn}},
  \bibinfo{journal}{Phys. Rev. Lett.} \textbf{\bibinfo{volume}{38}},
  \bibinfo{pages}{1440} (\bibinfo{year}{1977}{\natexlab{a}}).

\bibitem[{\citenamefont{Peccei and Quinn}(1977{\natexlab{b}})}]{Peccei:1977ur}
\bibinfo{author}{\bibfnamefont{R.~D.} \bibnamefont{Peccei}} \bibnamefont{and}
  \bibinfo{author}{\bibfnamefont{H.~R.} \bibnamefont{Quinn}},
  \bibinfo{journal}{Phys. Rev.} \textbf{\bibinfo{volume}{D16}},
  \bibinfo{pages}{1791} (\bibinfo{year}{1977}{\natexlab{b}}).

\bibitem[{\citenamefont{Weinberg}(1978)}]{Weinberg:1977ma}
\bibinfo{author}{\bibfnamefont{S.}~\bibnamefont{Weinberg}},
  \bibinfo{journal}{Phys. Rev. Lett.} \textbf{\bibinfo{volume}{40}},
  \bibinfo{pages}{223} (\bibinfo{year}{1978}).

\bibitem[{\citenamefont{Wilczek}(1978)}]{Wilczek:1977pj}
\bibinfo{author}{\bibfnamefont{F.}~\bibnamefont{Wilczek}},
  \bibinfo{journal}{Phys. Rev. Lett.} \textbf{\bibinfo{volume}{40}},
  \bibinfo{pages}{279} (\bibinfo{year}{1978}).

\bibitem[{\citenamefont{McKay}(1977)}]{McKay:1977gd}
\bibinfo{author}{\bibfnamefont{D.~W.} \bibnamefont{McKay}},
  \bibinfo{journal}{Phys. Rev.} \textbf{\bibinfo{volume}{D16}},
  \bibinfo{pages}{2861} (\bibinfo{year}{1977}).

\bibitem[{\citenamefont{McKay and Munczek}(1979)}]{McKay:1978wn}
\bibinfo{author}{\bibfnamefont{D.~W.} \bibnamefont{McKay}} \bibnamefont{and}
  \bibinfo{author}{\bibfnamefont{H.}~\bibnamefont{Munczek}},
  \bibinfo{journal}{Phys. Rev.} \textbf{\bibinfo{volume}{D19}},
  \bibinfo{pages}{985} (\bibinfo{year}{1979}).

\bibitem[{\citenamefont{Kim}(1979)}]{Kim:1979if}
\bibinfo{author}{\bibfnamefont{J.~E.} \bibnamefont{Kim}},
  \bibinfo{journal}{Phys. Rev. Lett.} \textbf{\bibinfo{volume}{43}},
  \bibinfo{pages}{103} (\bibinfo{year}{1979}).

\bibitem[{\citenamefont{Dine et~al.}(1981)\citenamefont{Dine, Fischler, and
  Srednicki}}]{Dine:1981rt}
\bibinfo{author}{\bibfnamefont{M.}~\bibnamefont{Dine}},
  \bibinfo{author}{\bibfnamefont{W.}~\bibnamefont{Fischler}}, \bibnamefont{and}
  \bibinfo{author}{\bibfnamefont{M.}~\bibnamefont{Srednicki}},
  \bibinfo{journal}{Phys. Lett.} \textbf{\bibinfo{volume}{B104}},
  \bibinfo{pages}{199} (\bibinfo{year}{1981}).

\bibitem[{\citenamefont{Kim}(1987)}]{Kim:1986ax}
\bibinfo{author}{\bibfnamefont{J.~E.} \bibnamefont{Kim}},
  \bibinfo{journal}{Phys. Rept.} \textbf{\bibinfo{volume}{150}},
  \bibinfo{pages}{1} (\bibinfo{year}{1987}).

\bibitem[{\citenamefont{Clarke et~al.}(1982)\citenamefont{Clarke, Karl, and
  Watson}}]{Clarke:1982}
\bibinfo{author}{\bibfnamefont{J.~N.} \bibnamefont{Clarke}},
  \bibinfo{author}{\bibfnamefont{G.}~\bibnamefont{Karl}}, \bibnamefont{and}
  \bibinfo{author}{\bibfnamefont{P.~J.~S.} \bibnamefont{Watson}},
  \bibinfo{journal}{Can. J. Phys.} \textbf{\bibinfo{volume}{60}},
  \bibinfo{pages}{1561} (\bibinfo{year}{1982}).

\bibitem[{\citenamefont{Sikivie}(1983)}]{Sikivie:1983ip}
\bibinfo{author}{\bibfnamefont{P.}~\bibnamefont{Sikivie}},
  \bibinfo{journal}{Phys. Rev. Lett.} \textbf{\bibinfo{volume}{51}},
  \bibinfo{pages}{1415} (\bibinfo{year}{1983}).

\bibitem[{\citenamefont{Sikivie}(1985)}]{Sikivie:1985yu}
\bibinfo{author}{\bibfnamefont{P.}~\bibnamefont{Sikivie}},
  \bibinfo{journal}{Phys. Rev.} \textbf{\bibinfo{volume}{D32}},
  \bibinfo{pages}{2988} (\bibinfo{year}{1985}).

\bibitem[{\citenamefont{Sikivie}(1988)}]{Sikivie:1988mz}
\bibinfo{author}{\bibfnamefont{P.}~\bibnamefont{Sikivie}},
  \bibinfo{journal}{Phys. Rev. Lett.} \textbf{\bibinfo{volume}{61}},
  \bibinfo{pages}{783} (\bibinfo{year}{1988}).

\bibitem[{\citenamefont{Maiani et~al.}(1986)\citenamefont{Maiani, Petronzio,
  and Zavattini}}]{Maiani:1986md}
\bibinfo{author}{\bibfnamefont{L.}~\bibnamefont{Maiani}},
  \bibinfo{author}{\bibfnamefont{R.}~\bibnamefont{Petronzio}},
  \bibnamefont{and}
  \bibinfo{author}{\bibfnamefont{E.}~\bibnamefont{Zavattini}},
  \bibinfo{journal}{Phys. Lett.} \textbf{\bibinfo{volume}{B175}},
  \bibinfo{pages}{359} (\bibinfo{year}{1986}).

\bibitem[{\citenamefont{Raffelt and Stodolsky}(1988)}]{Raffelt:1987im}
\bibinfo{author}{\bibfnamefont{G.}~\bibnamefont{Raffelt}} \bibnamefont{and}
  \bibinfo{author}{\bibfnamefont{L.}~\bibnamefont{Stodolsky}},
  \bibinfo{journal}{Phys. Rev.} \textbf{\bibinfo{volume}{D37}},
  \bibinfo{pages}{1237} (\bibinfo{year}{1988}).

\bibitem[{\citenamefont{Carlson and Garretson}(1994)}]{Carlson:1994yqa}
\bibinfo{author}{\bibfnamefont{E.~D.} \bibnamefont{Carlson}} \bibnamefont{and}
  \bibinfo{author}{\bibfnamefont{W.~D.} \bibnamefont{Garretson}},
  \bibinfo{journal}{Phys. Lett.} \textbf{\bibinfo{volume}{B336}},
  \bibinfo{pages}{431} (\bibinfo{year}{1994}).

\bibitem[{\citenamefont{Bradley et~al.}(2003)}]{Bradley:2003kg}
\bibinfo{author}{\bibfnamefont{R.}~\bibnamefont{Bradley}} \bibnamefont{et~al.},
  \bibinfo{journal}{Rev. Mod. Phys.} \textbf{\bibinfo{volume}{75}},
  \bibinfo{pages}{777} (\bibinfo{year}{2003}).

\bibitem[{\citenamefont{Das et~al.}(2005)\citenamefont{Das, Jain, Ralston, and
  Saha}}]{Das:2004qka}
\bibinfo{author}{\bibfnamefont{S.}~\bibnamefont{Das}},
  \bibinfo{author}{\bibfnamefont{P.}~\bibnamefont{Jain}},
  \bibinfo{author}{\bibfnamefont{J.~P.} \bibnamefont{Ralston}},
  \bibnamefont{and} \bibinfo{author}{\bibfnamefont{R.}~\bibnamefont{Saha}},
  \bibinfo{journal}{JCAP} \textbf{\bibinfo{volume}{0506}}, \bibinfo{pages}{002}
  (\bibinfo{year}{2005}), \eprint{hep-ph/0408198}.

\bibitem[{\citenamefont{Das et~al.}(2008)\citenamefont{Das, Jain, Ralston, and
  Saha}}]{Das:2004ee}
\bibinfo{author}{\bibfnamefont{S.}~\bibnamefont{Das}},
  \bibinfo{author}{\bibfnamefont{P.}~\bibnamefont{Jain}},
  \bibinfo{author}{\bibfnamefont{J.~P.} \bibnamefont{Ralston}},
  \bibnamefont{and} \bibinfo{author}{\bibfnamefont{R.}~\bibnamefont{Saha}},
  \bibinfo{journal}{Pramana} \textbf{\bibinfo{volume}{70}},
  \bibinfo{pages}{439} (\bibinfo{year}{2008}), \eprint{hep-ph/0410006}.

\bibitem[{\citenamefont{Ganguly}(2006)}]{Ganguly:2005se}
\bibinfo{author}{\bibfnamefont{A.~K.} \bibnamefont{Ganguly}},
  \bibinfo{journal}{Annals Phys.} \textbf{\bibinfo{volume}{321}},
  \bibinfo{pages}{1457} (\bibinfo{year}{2006}), \eprint{hep-ph/0512323}.

\bibitem[{\citenamefont{Ganguly et~al.}(2009)\citenamefont{Ganguly, Jain, and
  Mandal}}]{Ganguly:2008kh}
\bibinfo{author}{\bibfnamefont{A.~K.} \bibnamefont{Ganguly}},
  \bibinfo{author}{\bibfnamefont{P.}~\bibnamefont{Jain}}, \bibnamefont{and}
  \bibinfo{author}{\bibfnamefont{S.}~\bibnamefont{Mandal}},
  \bibinfo{journal}{Phys. Rev.} \textbf{\bibinfo{volume}{D79}},
  \bibinfo{pages}{115014} (\bibinfo{year}{2009}), \eprint{0810.4380}.

\bibitem[{\citenamefont{Harari and Sikivie}(1992)}]{Harari:1992ea}
\bibinfo{author}{\bibfnamefont{D.}~\bibnamefont{Harari}} \bibnamefont{and}
  \bibinfo{author}{\bibfnamefont{P.}~\bibnamefont{Sikivie}},
  \bibinfo{journal}{Phys. Lett.} \textbf{\bibinfo{volume}{B289}},
  \bibinfo{pages}{67} (\bibinfo{year}{1992}).

\bibitem[{\citenamefont{Berezhiani et~al.}(1992)\citenamefont{Berezhiani,
  Sakharov, and Khlopov}}]{Berezhiani}
\bibinfo{author}{\bibfnamefont{Z.~G.} \bibnamefont{Berezhiani}},
  \bibinfo{author}{\bibfnamefont{A.~S.} \bibnamefont{Sakharov}},
  \bibnamefont{and} \bibinfo{author}{\bibfnamefont{M.~Y.}
  \bibnamefont{Khlopov}}, \bibinfo{journal}{Yadernaya Fizika \bf{55}, 1918
  [Sov. J. Nucl. Phys. \bf{55}, 1063]}  (\bibinfo{year}{1992}).

\bibitem[{\citenamefont{Mohanty and Nayak}(1993)}]{Mohanty:1993nh}
\bibinfo{author}{\bibfnamefont{S.}~\bibnamefont{Mohanty}} \bibnamefont{and}
  \bibinfo{author}{\bibfnamefont{S.~N.} \bibnamefont{Nayak}},
  \bibinfo{journal}{Phys. Rev. Lett.} \textbf{\bibinfo{volume}{70}},
  \bibinfo{pages}{4038} (\bibinfo{year}{1993}), \eprint{hep-ph/9303310}.

\bibitem[{\citenamefont{Das et~al.}(2001)\citenamefont{Das, Jain, and
  Mukherji}}]{Das:2000ph}
\bibinfo{author}{\bibfnamefont{P.}~\bibnamefont{Das}},
  \bibinfo{author}{\bibfnamefont{P.}~\bibnamefont{Jain}}, \bibnamefont{and}
  \bibinfo{author}{\bibfnamefont{S.}~\bibnamefont{Mukherji}},
  \bibinfo{journal}{Int. J. Mod. Phys.} \textbf{\bibinfo{volume}{A16}},
  \bibinfo{pages}{4011} (\bibinfo{year}{2001}), \eprint{hep-ph/0011279}.

\bibitem[{\citenamefont{Kar et~al.}(2002{\natexlab{a}})\citenamefont{Kar,
  Majumdar, SenGupta, and Sinha}}]{Kar:2000ct}
\bibinfo{author}{\bibfnamefont{S.}~\bibnamefont{Kar}},
  \bibinfo{author}{\bibfnamefont{P.}~\bibnamefont{Majumdar}},
  \bibinfo{author}{\bibfnamefont{S.}~\bibnamefont{SenGupta}}, \bibnamefont{and}
  \bibinfo{author}{\bibfnamefont{A.}~\bibnamefont{Sinha}},
  \bibinfo{journal}{Eur. Phys. J.} \textbf{\bibinfo{volume}{C23}},
  \bibinfo{pages}{357} (\bibinfo{year}{2002}{\natexlab{a}}),
  \eprint{gr-qc/0006097}.

\bibitem[{\citenamefont{Kar et~al.}(2002{\natexlab{b}})\citenamefont{Kar,
  Majumdar, SenGupta, and Sur}}]{Kar:2001eb}
\bibinfo{author}{\bibfnamefont{S.}~\bibnamefont{Kar}},
  \bibinfo{author}{\bibfnamefont{P.}~\bibnamefont{Majumdar}},
  \bibinfo{author}{\bibfnamefont{S.}~\bibnamefont{SenGupta}}, \bibnamefont{and}
  \bibinfo{author}{\bibfnamefont{S.}~\bibnamefont{Sur}},
  \bibinfo{journal}{Class. Quant. Grav.} \textbf{\bibinfo{volume}{19}},
  \bibinfo{pages}{677} (\bibinfo{year}{2002}{\natexlab{b}}),
  \eprint{hep-th/0109135}.

\bibitem[{\citenamefont{Csaki et~al.}(2002{\natexlab{a}})\citenamefont{Csaki,
  Kaloper, and Terning}}]{Csaki:2001jk}
\bibinfo{author}{\bibfnamefont{C.}~\bibnamefont{Csaki}},
  \bibinfo{author}{\bibfnamefont{N.}~\bibnamefont{Kaloper}}, \bibnamefont{and}
  \bibinfo{author}{\bibfnamefont{J.}~\bibnamefont{Terning}},
  \bibinfo{journal}{Phys. Lett.} \textbf{\bibinfo{volume}{B535}},
  \bibinfo{pages}{33} (\bibinfo{year}{2002}{\natexlab{a}}),
  \eprint{hep-ph/0112212}.

\bibitem[{\citenamefont{Csaki et~al.}(2002{\natexlab{b}})\citenamefont{Csaki,
  Kaloper, and Terning}}]{Csaki:2001yk}
\bibinfo{author}{\bibfnamefont{C.}~\bibnamefont{Csaki}},
  \bibinfo{author}{\bibfnamefont{N.}~\bibnamefont{Kaloper}}, \bibnamefont{and}
  \bibinfo{author}{\bibfnamefont{J.}~\bibnamefont{Terning}},
  \bibinfo{journal}{Phys. Rev. Lett.} \textbf{\bibinfo{volume}{88}},
  \bibinfo{pages}{161302} (\bibinfo{year}{2002}{\natexlab{b}}),
  \eprint{hep-ph/0111311}.

\bibitem[{\citenamefont{Grossman et~al.}(2002)\citenamefont{Grossman, Roy, and
  Zupan}}]{Grossman:2002by}
\bibinfo{author}{\bibfnamefont{Y.}~\bibnamefont{Grossman}},
  \bibinfo{author}{\bibfnamefont{S.}~\bibnamefont{Roy}}, \bibnamefont{and}
  \bibinfo{author}{\bibfnamefont{J.}~\bibnamefont{Zupan}},
  \bibinfo{journal}{Phys. Lett.} \textbf{\bibinfo{volume}{B543}},
  \bibinfo{pages}{23} (\bibinfo{year}{2002}), \eprint{hep-ph/0204216}.

\bibitem[{\citenamefont{Jain et~al.}(2002)\citenamefont{Jain, Panda, and
  Sarala}}]{Jain:2002vx}
\bibinfo{author}{\bibfnamefont{P.}~\bibnamefont{Jain}},
  \bibinfo{author}{\bibfnamefont{S.}~\bibnamefont{Panda}}, \bibnamefont{and}
  \bibinfo{author}{\bibfnamefont{S.}~\bibnamefont{Sarala}},
  \bibinfo{journal}{Phys. Rev.} \textbf{\bibinfo{volume}{D66}},
  \bibinfo{pages}{085007} (\bibinfo{year}{2002}), \eprint{hep-ph/0206046}.

\bibitem[{\citenamefont{Song and Hu}(2006)}]{Song:2005af}
\bibinfo{author}{\bibfnamefont{Y.-S.} \bibnamefont{Song}} \bibnamefont{and}
  \bibinfo{author}{\bibfnamefont{W.}~\bibnamefont{Hu}}, \bibinfo{journal}{Phys.
  Rev.} \textbf{\bibinfo{volume}{D73}}, \bibinfo{pages}{023003}
  (\bibinfo{year}{2006}), \eprint{astro-ph/0508002}.

\bibitem[{\citenamefont{Mirizzi et~al.}(2005)\citenamefont{Mirizzi, Raffelt,
  and Serpico}}]{Mirizzi:2005ng}
\bibinfo{author}{\bibfnamefont{A.}~\bibnamefont{Mirizzi}},
  \bibinfo{author}{\bibfnamefont{G.~G.} \bibnamefont{Raffelt}},
  \bibnamefont{and} \bibinfo{author}{\bibfnamefont{P.~D.}
  \bibnamefont{Serpico}}, \bibinfo{journal}{Phys. Rev.}
  \textbf{\bibinfo{volume}{D72}}, \bibinfo{pages}{023501}
  (\bibinfo{year}{2005}), \eprint{astro-ph/0506078}.

\bibitem[{\citenamefont{Raffelt}(2008)}]{Raffelt:2006cw}
\bibinfo{author}{\bibfnamefont{G.~G.} \bibnamefont{Raffelt}},
  \bibinfo{journal}{Lect. Notes Phys.} \textbf{\bibinfo{volume}{741}},
  \bibinfo{pages}{51} (\bibinfo{year}{2008}), \eprint{hep-ph/0611350}.

\bibitem[{\citenamefont{Gnedin et~al.}(2007)\citenamefont{Gnedin, Piotrovich,
  and Natsvlishvili}}]{Gnedin:2006fq}
\bibinfo{author}{\bibfnamefont{Y.~N.} \bibnamefont{Gnedin}},
  \bibinfo{author}{\bibfnamefont{M.~Y.} \bibnamefont{Piotrovich}},
  \bibnamefont{and} \bibinfo{author}{\bibfnamefont{T.~M.}
  \bibnamefont{Natsvlishvili}}, \bibinfo{journal}{Mon. Not. Roy. Astron. Soc.}
  \textbf{\bibinfo{volume}{374}}, \bibinfo{pages}{276} (\bibinfo{year}{2007}),
  \eprint{astro-ph/0607294}.

\bibitem[{\citenamefont{Mirizzi et~al.}(2007)\citenamefont{Mirizzi, Raffelt,
  and Serpico}}]{Mirizzi:2007hr}
\bibinfo{author}{\bibfnamefont{A.}~\bibnamefont{Mirizzi}},
  \bibinfo{author}{\bibfnamefont{G.~G.} \bibnamefont{Raffelt}},
  \bibnamefont{and} \bibinfo{author}{\bibfnamefont{P.~D.}
  \bibnamefont{Serpico}}, \bibinfo{journal}{Phys. Rev.}
  \textbf{\bibinfo{volume}{D76}}, \bibinfo{pages}{023001}
  (\bibinfo{year}{2007}), \eprint{0704.3044}.

\bibitem[{\citenamefont{Finelli and Galaverni}(2009)}]{Finelli:2008jv}
\bibinfo{author}{\bibfnamefont{F.}~\bibnamefont{Finelli}} \bibnamefont{and}
  \bibinfo{author}{\bibfnamefont{M.}~\bibnamefont{Galaverni}},
  \bibinfo{journal}{Phys. Rev.} \textbf{\bibinfo{volume}{D79}},
  \bibinfo{pages}{063002} (\bibinfo{year}{2009}), \eprint{0802.4210}.

\bibitem[{\citenamefont{Ostman and Mortsell}(2005)}]{Ostman:2004eh}
\bibinfo{author}{\bibfnamefont{L.}~\bibnamefont{Ostman}} \bibnamefont{and}
  \bibinfo{author}{\bibfnamefont{E.}~\bibnamefont{Mortsell}},
  \bibinfo{journal}{JCAP} \textbf{\bibinfo{volume}{0502}}, \bibinfo{pages}{005}
  (\bibinfo{year}{2005}), \eprint{astro-ph/0410501}.

\bibitem[{\citenamefont{Lai and Heyl}(2006)}]{Lai:2006af}
\bibinfo{author}{\bibfnamefont{D.}~\bibnamefont{Lai}} \bibnamefont{and}
  \bibinfo{author}{\bibfnamefont{J.}~\bibnamefont{Heyl}},
  \bibinfo{journal}{Phys. Rev.} \textbf{\bibinfo{volume}{D74}},
  \bibinfo{pages}{123003} (\bibinfo{year}{2006}), \eprint{astro-ph/0609775}.

\bibitem[{\citenamefont{Hooper and Serpico}(2007)}]{Hooper:2007bq}
\bibinfo{author}{\bibfnamefont{D.}~\bibnamefont{Hooper}} \bibnamefont{and}
  \bibinfo{author}{\bibfnamefont{P.~D.} \bibnamefont{Serpico}},
  \bibinfo{journal}{Phys. Rev. Lett.} \textbf{\bibinfo{volume}{99}},
  \bibinfo{pages}{231102} (\bibinfo{year}{2007}), \eprint{0706.3203}.

\bibitem[{\citenamefont{Hochmuth and Sigl}(2007)}]{Hochmuth:2007hk}
\bibinfo{author}{\bibfnamefont{K.~A.} \bibnamefont{Hochmuth}} \bibnamefont{and}
  \bibinfo{author}{\bibfnamefont{G.}~\bibnamefont{Sigl}},
  \bibinfo{journal}{Phys. Rev.} \textbf{\bibinfo{volume}{D76}},
  \bibinfo{pages}{123011} (\bibinfo{year}{2007}), \eprint{0708.1144}.

\bibitem[{\citenamefont{Chelouche et~al.}(2009)\citenamefont{Chelouche,
  Rabadan, Pavlov, and Castejon}}]{Chelouche:2008ta}
\bibinfo{author}{\bibfnamefont{D.}~\bibnamefont{Chelouche}},
  \bibinfo{author}{\bibfnamefont{R.}~\bibnamefont{Rabadan}},
  \bibinfo{author}{\bibfnamefont{S.}~\bibnamefont{Pavlov}}, \bibnamefont{and}
  \bibinfo{author}{\bibfnamefont{F.}~\bibnamefont{Castejon}},
  \bibinfo{journal}{Astrophys. J. Suppl.} \textbf{\bibinfo{volume}{180}},
  \bibinfo{pages}{1} (\bibinfo{year}{2009}), \eprint{0806.0411}.

\bibitem[{\citenamefont{Dicus et~al.}(1978)\citenamefont{Dicus, Kolb, Teplitz,
  and Wagoner}}]{Dicus:1978fp}
\bibinfo{author}{\bibfnamefont{D.~A.} \bibnamefont{Dicus}},
  \bibinfo{author}{\bibfnamefont{E.~W.} \bibnamefont{Kolb}},
  \bibinfo{author}{\bibfnamefont{V.~L.} \bibnamefont{Teplitz}},
  \bibnamefont{and} \bibinfo{author}{\bibfnamefont{R.~V.}
  \bibnamefont{Wagoner}}, \bibinfo{journal}{Phys. Rev.}
  \textbf{\bibinfo{volume}{D18}}, \bibinfo{pages}{1829} (\bibinfo{year}{1978}).

\bibitem[{\citenamefont{Vysotsskii et~al.}(1978)\citenamefont{Vysotsskii,
  Zel'Dovich, Khlopov, and Chechetkin}}]{Vysotsskii:1978}
\bibinfo{author}{\bibfnamefont{M.~I.} \bibnamefont{Vysotsskii}},
  \bibinfo{author}{\bibfnamefont{Y.~B.} \bibnamefont{Zel'Dovich}},
  \bibinfo{author}{\bibfnamefont{M.~Y.} \bibnamefont{Khlopov}},
  \bibnamefont{and} \bibinfo{author}{\bibfnamefont{V.~M.}
  \bibnamefont{Chechetkin}}, \bibinfo{journal}{Pis'ma ZhETF \bf{27}, 533 [JETP
  Lett. \bf{27}, 502]}  (\bibinfo{year}{1978}).

\bibitem[{\citenamefont{Dearborn et~al.}(1986)\citenamefont{Dearborn, Schramm,
  and Steigman}}]{Dearborn:1985gp}
\bibinfo{author}{\bibfnamefont{D.~S.~P.} \bibnamefont{Dearborn}},
  \bibinfo{author}{\bibfnamefont{D.~N.} \bibnamefont{Schramm}},
  \bibnamefont{and} \bibinfo{author}{\bibfnamefont{G.}~\bibnamefont{Steigman}},
  \bibinfo{journal}{Phys. Rev. Lett.} \textbf{\bibinfo{volume}{56}},
  \bibinfo{pages}{26} (\bibinfo{year}{1986}).

\bibitem[{\citenamefont{Raffelt and Dearborn}(1987)}]{Raffelt:1987yu}
\bibinfo{author}{\bibfnamefont{G.~G.} \bibnamefont{Raffelt}} \bibnamefont{and}
  \bibinfo{author}{\bibfnamefont{D.~S.~P.} \bibnamefont{Dearborn}},
  \bibinfo{journal}{Phys. Rev.} \textbf{\bibinfo{volume}{D36}},
  \bibinfo{pages}{2211} (\bibinfo{year}{1987}).

\bibitem[{\citenamefont{Raffelt and Seckel}(1988)}]{Raffelt:1987yt}
\bibinfo{author}{\bibfnamefont{G.}~\bibnamefont{Raffelt}} \bibnamefont{and}
  \bibinfo{author}{\bibfnamefont{D.}~\bibnamefont{Seckel}},
  \bibinfo{journal}{Phys. Rev. Lett.} \textbf{\bibinfo{volume}{60}},
  \bibinfo{pages}{1793} (\bibinfo{year}{1988}).

\bibitem[{\citenamefont{Turner}(1988)}]{Turner:1987by}
\bibinfo{author}{\bibfnamefont{M.~S.} \bibnamefont{Turner}},
  \bibinfo{journal}{Phys. Rev. Lett.} \textbf{\bibinfo{volume}{60}},
  \bibinfo{pages}{1797} (\bibinfo{year}{1988}).

\bibitem[{\citenamefont{Janka et~al.}(1996)\citenamefont{Janka, Keil, Raffelt,
  and Seckel}}]{Janka:1995ir}
\bibinfo{author}{\bibfnamefont{H.-T.} \bibnamefont{Janka}},
  \bibinfo{author}{\bibfnamefont{W.}~\bibnamefont{Keil}},
  \bibinfo{author}{\bibfnamefont{G.}~\bibnamefont{Raffelt}}, \bibnamefont{and}
  \bibinfo{author}{\bibfnamefont{D.}~\bibnamefont{Seckel}},
  \bibinfo{journal}{Phys. Rev. Lett.} \textbf{\bibinfo{volume}{76}},
  \bibinfo{pages}{2621} (\bibinfo{year}{1996}), \eprint{astro-ph/9507023}.

\bibitem[{\citenamefont{Keil et~al.}(1997)}]{Keil:1996ju}
\bibinfo{author}{\bibfnamefont{W.}~\bibnamefont{Keil}} \bibnamefont{et~al.},
  \bibinfo{journal}{Phys. Rev.} \textbf{\bibinfo{volume}{D56}},
  \bibinfo{pages}{2419} (\bibinfo{year}{1997}), \eprint{astro-ph/9612222}.

\bibitem[{\citenamefont{Brockway et~al.}(1996)\citenamefont{Brockway, Carlson,
  and Raffelt}}]{Brockway:1996yr}
\bibinfo{author}{\bibfnamefont{J.~W.} \bibnamefont{Brockway}},
  \bibinfo{author}{\bibfnamefont{E.~D.} \bibnamefont{Carlson}},
  \bibnamefont{and} \bibinfo{author}{\bibfnamefont{G.~G.}
  \bibnamefont{Raffelt}}, \bibinfo{journal}{Phys. Lett.}
  \textbf{\bibinfo{volume}{B383}}, \bibinfo{pages}{439} (\bibinfo{year}{1996}),
  \eprint{astro-ph/9605197}.

\bibitem[{\citenamefont{Grifols et~al.}(1996)\citenamefont{Grifols, Masso, and
  Toldra}}]{Grifols:1996id}
\bibinfo{author}{\bibfnamefont{J.~A.} \bibnamefont{Grifols}},
  \bibinfo{author}{\bibfnamefont{E.}~\bibnamefont{Masso}}, \bibnamefont{and}
  \bibinfo{author}{\bibfnamefont{R.}~\bibnamefont{Toldra}},
  \bibinfo{journal}{Phys. Rev. Lett.} \textbf{\bibinfo{volume}{77}},
  \bibinfo{pages}{2372} (\bibinfo{year}{1996}), \eprint{astro-ph/9606028}.

\bibitem[{\citenamefont{Raffelt}(1999)}]{Raffelt:1999tx}
\bibinfo{author}{\bibfnamefont{G.~G.} \bibnamefont{Raffelt}},
  \bibinfo{journal}{Ann. Rev. Nucl. Part. Sci.} \textbf{\bibinfo{volume}{49}},
  \bibinfo{pages}{163} (\bibinfo{year}{1999}), \eprint{hep-ph/9903472}.

\bibitem[{\citenamefont{Rosenberg and van Bibber}(2000)}]{Rosenberg:2000wb}
\bibinfo{author}{\bibfnamefont{L.~J.} \bibnamefont{Rosenberg}}
  \bibnamefont{and} \bibinfo{author}{\bibfnamefont{K.~A.} \bibnamefont{van
  Bibber}}, \bibinfo{journal}{Phys. Rept.} \textbf{\bibinfo{volume}{325}},
  \bibinfo{pages}{1} (\bibinfo{year}{2000}).

\bibitem[{\citenamefont{Zioutas et~al.}(2005)}]{Zioutas:2004hi}
\bibinfo{author}{\bibfnamefont{K.}~\bibnamefont{Zioutas}} \bibnamefont{et~al.}
  (\bibinfo{collaboration}{CAST}), \bibinfo{journal}{Phys. Rev. Lett.}
  \textbf{\bibinfo{volume}{94}}, \bibinfo{pages}{121301}
  (\bibinfo{year}{2005}), \eprint{hep-ex/0411033}.

\bibitem[{\citenamefont{Yao et~al.}(2006)}]{Yao:2006px}
\bibinfo{author}{\bibfnamefont{W.~M.} \bibnamefont{Yao}} \bibnamefont{et~al.}
  (\bibinfo{collaboration}{Particle Data Group}), \bibinfo{journal}{J. Phys.}
  \textbf{\bibinfo{volume}{G33}}, \bibinfo{pages}{1} (\bibinfo{year}{2006}).

\bibitem[{\citenamefont{Jaeckel et~al.}(2007)\citenamefont{Jaeckel, Masso,
  Redondo, Ringwald, and Takahashi}}]{Jaeckel:2006xm}
\bibinfo{author}{\bibfnamefont{J.}~\bibnamefont{Jaeckel}},
  \bibinfo{author}{\bibfnamefont{E.}~\bibnamefont{Masso}},
  \bibinfo{author}{\bibfnamefont{J.}~\bibnamefont{Redondo}},
  \bibinfo{author}{\bibfnamefont{A.}~\bibnamefont{Ringwald}}, \bibnamefont{and}
  \bibinfo{author}{\bibfnamefont{F.}~\bibnamefont{Takahashi}},
  \bibinfo{journal}{Phys. Rev.} \textbf{\bibinfo{volume}{D75}},
  \bibinfo{pages}{013004} (\bibinfo{year}{2007}), \eprint{hep-ph/0610203}.

\bibitem[{\citenamefont{Andriamonje et~al.}(2007)}]{Andriamonje:2007ew}
\bibinfo{author}{\bibfnamefont{S.}~\bibnamefont{Andriamonje}}
  \bibnamefont{et~al.} (\bibinfo{collaboration}{CAST}), \bibinfo{journal}{JCAP}
  \textbf{\bibinfo{volume}{0704}}, \bibinfo{pages}{010} (\bibinfo{year}{2007}),
  \eprint{hep-ex/0702006}.

\bibitem[{\citenamefont{Robilliard et~al.}(2007)}]{Robilliard:2007bq}
\bibinfo{author}{\bibfnamefont{C.}~\bibnamefont{Robilliard}}
  \bibnamefont{et~al.}, \bibinfo{journal}{Phys. Rev. Lett.}
  \textbf{\bibinfo{volume}{99}}, \bibinfo{pages}{190403}
  (\bibinfo{year}{2007}), \eprint{0707.1296}.

\bibitem[{\citenamefont{Zavattini et~al.}(2008)}]{Zavattini:2007ee}
\bibinfo{author}{\bibfnamefont{E.}~\bibnamefont{Zavattini}}
  \bibnamefont{et~al.} (\bibinfo{collaboration}{PVLAS}),
  \bibinfo{journal}{Phys. Rev.} \textbf{\bibinfo{volume}{D77}},
  \bibinfo{pages}{032006} (\bibinfo{year}{2008}), \eprint{0706.3419}.

\bibitem[{\citenamefont{Rubbia and Sakharov}(2008)}]{Rubbia:2007hf}
\bibinfo{author}{\bibfnamefont{A.}~\bibnamefont{Rubbia}} \bibnamefont{and}
  \bibinfo{author}{\bibfnamefont{A.~S.} \bibnamefont{Sakharov}},
  \bibinfo{journal}{Astropart. Phys.} \textbf{\bibinfo{volume}{29}},
  \bibinfo{pages}{20} (\bibinfo{year}{2008}), \eprint{0708.2646}.

\bibitem[{\citenamefont{Hutsemekers}(1998)}]{Hutsemekers:1998}
\bibinfo{author}{\bibfnamefont{D.}~\bibnamefont{Hutsemekers}},
  \bibinfo{journal}{Astron. Astrophys.} \textbf{\bibinfo{volume}{332}},
  \bibinfo{pages}{410} (\bibinfo{year}{1998}).

\bibitem[{\citenamefont{Hutsemekers and Lamy}(2001)}]{Hutsemekers:2000fv}
\bibinfo{author}{\bibfnamefont{D.}~\bibnamefont{Hutsemekers}} \bibnamefont{and}
  \bibinfo{author}{\bibfnamefont{H.}~\bibnamefont{Lamy}},
  \bibinfo{journal}{Astron. Astrophys.} \textbf{\bibinfo{volume}{367}},
  \bibinfo{pages}{381} (\bibinfo{year}{2001}), \eprint{astro-ph/0012182}.

\bibitem[{\citenamefont{Hutsemekers et~al.}(2005)\citenamefont{Hutsemekers,
  Cabanac, Lamy, and Sluse}}]{Hutsemekers:2005iz}
\bibinfo{author}{\bibfnamefont{D.}~\bibnamefont{Hutsemekers}},
  \bibinfo{author}{\bibfnamefont{R.}~\bibnamefont{Cabanac}},
  \bibinfo{author}{\bibfnamefont{H.}~\bibnamefont{Lamy}}, \bibnamefont{and}
  \bibinfo{author}{\bibfnamefont{D.}~\bibnamefont{Sluse}},
  \bibinfo{journal}{Astron. Astrophys.} \textbf{\bibinfo{volume}{441}},
  \bibinfo{pages}{915} (\bibinfo{year}{2005}), \eprint{astro-ph/0507274}.

\bibitem[{\citenamefont{Jain and Ralston}(1999)}]{Jain:1998kf}
\bibinfo{author}{\bibfnamefont{P.}~\bibnamefont{Jain}} \bibnamefont{and}
  \bibinfo{author}{\bibfnamefont{J.~P.} \bibnamefont{Ralston}},
  \bibinfo{journal}{Mod. Phys. Lett.} \textbf{\bibinfo{volume}{A14}},
  \bibinfo{pages}{417} (\bibinfo{year}{1999}), \eprint{astro-ph/9803164}.

\bibitem[{\citenamefont{Ralston and Jain}(2004)}]{Ralston:2003pf}
\bibinfo{author}{\bibfnamefont{J.~P.} \bibnamefont{Ralston}} \bibnamefont{and}
  \bibinfo{author}{\bibfnamefont{P.}~\bibnamefont{Jain}},
  \bibinfo{journal}{Int. J. Mod. Phys.} \textbf{\bibinfo{volume}{D13}},
  \bibinfo{pages}{1857} (\bibinfo{year}{2004}), \eprint{astro-ph/0311430}.

\bibitem[{\citenamefont{Jain et~al.}(2004)\citenamefont{Jain, Narain, and
  Sarala}}]{Jain:2003sg}
\bibinfo{author}{\bibfnamefont{P.}~\bibnamefont{Jain}},
  \bibinfo{author}{\bibfnamefont{G.}~\bibnamefont{Narain}}, \bibnamefont{and}
  \bibinfo{author}{\bibfnamefont{S.}~\bibnamefont{Sarala}},
  \bibinfo{journal}{Mon. Not. Roy. Astron. Soc.}
  \textbf{\bibinfo{volume}{347}}, \bibinfo{pages}{394} (\bibinfo{year}{2004}),
  \eprint{astro-ph/0301530}.

\bibitem[{\citenamefont{Payez et~al.}(2008)\citenamefont{Payez, Cudell, and
  Hutsemekers}}]{Payez:2008pm}
\bibinfo{author}{\bibfnamefont{A.}~\bibnamefont{Payez}},
  \bibinfo{author}{\bibfnamefont{J.~R.} \bibnamefont{Cudell}},
  \bibnamefont{and}
  \bibinfo{author}{\bibfnamefont{D.}~\bibnamefont{Hutsemekers}},
  \bibinfo{journal}{AIP Conf. Proc.} \textbf{\bibinfo{volume}{1038}},
  \bibinfo{pages}{211} (\bibinfo{year}{2008}), \eprint{0805.3946}.

\bibitem[{\citenamefont{Piotrovich et~al.}(2008)\citenamefont{Piotrovich,
  Gnedin, and Natsvlishvili}}]{Piotrovich:2008iy}
\bibinfo{author}{\bibfnamefont{M.~Y.} \bibnamefont{Piotrovich}},
  \bibinfo{author}{\bibfnamefont{Y.~N.} \bibnamefont{Gnedin}},
  \bibnamefont{and} \bibinfo{author}{\bibfnamefont{T.~M.}
  \bibnamefont{Natsvlishvili}} (\bibinfo{year}{2008}), \eprint{0805.3649}.

\bibitem[{\citenamefont{Lee et~al.}(2006)\citenamefont{Lee, Liu, and
  Ng}}]{Lee:2006za}
\bibinfo{author}{\bibfnamefont{S.}~\bibnamefont{Lee}},
  \bibinfo{author}{\bibfnamefont{G.-C.} \bibnamefont{Liu}}, \bibnamefont{and}
  \bibinfo{author}{\bibfnamefont{K.-W.} \bibnamefont{Ng}},
  \bibinfo{journal}{Phys. Rev.} \textbf{\bibinfo{volume}{D73}},
  \bibinfo{pages}{083516} (\bibinfo{year}{2006}), \eprint{astro-ph/0601333}.

\bibitem[{\citenamefont{Agarwal et~al.}(2008)\citenamefont{Agarwal, Jain,
  McKay, and Ralston}}]{Agarwal:2008ac}
\bibinfo{author}{\bibfnamefont{N.}~\bibnamefont{Agarwal}},
  \bibinfo{author}{\bibfnamefont{P.}~\bibnamefont{Jain}},
  \bibinfo{author}{\bibfnamefont{D.~W.} \bibnamefont{McKay}}, \bibnamefont{and}
  \bibinfo{author}{\bibfnamefont{J.~P.} \bibnamefont{Ralston}},
  \bibinfo{journal}{Phys. Rev.} \textbf{\bibinfo{volume}{D78}},
  \bibinfo{pages}{085028} (\bibinfo{year}{2008}), \eprint{0807.4587}.

\bibitem[{\citenamefont{Subramanian et~al.}(2003)\citenamefont{Subramanian,
  Seshadri, and Barrow}}]{Subramanian:2003sh}
\bibinfo{author}{\bibfnamefont{K.}~\bibnamefont{Subramanian}},
  \bibinfo{author}{\bibfnamefont{T.~R.} \bibnamefont{Seshadri}},
  \bibnamefont{and} \bibinfo{author}{\bibfnamefont{J.~D.}
  \bibnamefont{Barrow}}, \bibinfo{journal}{Mon. Not. Roy. Astron. Soc.}
  \textbf{\bibinfo{volume}{344}}, \bibinfo{pages}{L31} (\bibinfo{year}{2003}),
  \eprint{astro-ph/0303014}.

\bibitem[{\citenamefont{Seshadri and Subramanian}(2005)}]{Seshadri:2005aa}
\bibinfo{author}{\bibfnamefont{T.~R.} \bibnamefont{Seshadri}} \bibnamefont{and}
  \bibinfo{author}{\bibfnamefont{K.}~\bibnamefont{Subramanian}},
  \bibinfo{journal}{Phys. Rev.} \textbf{\bibinfo{volume}{D72}},
  \bibinfo{pages}{023004} (\bibinfo{year}{2005}), \eprint{astro-ph/0504007}.

\bibitem[{\citenamefont{Seshadri and Subramanian}(2009)}]{Seshadri:2009sy}
\bibinfo{author}{\bibfnamefont{T.~R.} \bibnamefont{Seshadri}} \bibnamefont{and}
  \bibinfo{author}{\bibfnamefont{K.}~\bibnamefont{Subramanian}},
  \bibinfo{journal}{Phys. Rev. Lett.} \textbf{\bibinfo{volume}{103}},
  \bibinfo{pages}{081303} (\bibinfo{year}{2009}), \eprint{0902.4066}.

\bibitem[{\citenamefont{Jedamzik et~al.}(1998)\citenamefont{Jedamzik,
  Katalinic, and Olinto}}]{Jedamzik:1996wp}
\bibinfo{author}{\bibfnamefont{K.}~\bibnamefont{Jedamzik}},
  \bibinfo{author}{\bibfnamefont{V.}~\bibnamefont{Katalinic}},
  \bibnamefont{and} \bibinfo{author}{\bibfnamefont{A.~V.}
  \bibnamefont{Olinto}}, \bibinfo{journal}{Phys. Rev.}
  \textbf{\bibinfo{volume}{D57}}, \bibinfo{pages}{3264} (\bibinfo{year}{1998}),
  \eprint{astro-ph/9606080}.

\bibitem[{\citenamefont{Subramanian and Barrow}(1998)}]{Subramanian:1997gi}
\bibinfo{author}{\bibfnamefont{K.}~\bibnamefont{Subramanian}} \bibnamefont{and}
  \bibinfo{author}{\bibfnamefont{J.~D.} \bibnamefont{Barrow}},
  \bibinfo{journal}{Phys. Rev.} \textbf{\bibinfo{volume}{D58}},
  \bibinfo{pages}{083502} (\bibinfo{year}{1998}), \eprint{astro-ph/9712083}.

\bibitem[{\citenamefont{Yamazaki et~al.}(2010)\citenamefont{Yamazaki, Ichiki,
  Kajino, and Mathews}}]{Yamazaki:2010nf}
\bibinfo{author}{\bibfnamefont{D.~G.} \bibnamefont{Yamazaki}},
  \bibinfo{author}{\bibfnamefont{K.}~\bibnamefont{Ichiki}},
  \bibinfo{author}{\bibfnamefont{T.}~\bibnamefont{Kajino}}, \bibnamefont{and}
  \bibinfo{author}{\bibfnamefont{G.~J.} \bibnamefont{Mathews}},
  \bibinfo{journal}{Phys. Rev.} \textbf{\bibinfo{volume}{D81}},
  \bibinfo{pages}{023008} (\bibinfo{year}{2010}), \eprint{1001.2012}.

\bibitem[{\citenamefont{Hutsemekers et~al.}(2008)}]{Hutsemekers:2008iv}
\bibinfo{author}{\bibfnamefont{D.}~\bibnamefont{Hutsemekers}}
  \bibnamefont{et~al.} (\bibinfo{year}{2008}), \eprint{0809.3088}.

\end{thebibliography}

\end{document}